\documentstyle[preprint,aps]{revtex}
\begin{document}
\newcommand{\be}{\begin{equation}}
\newcommand{\ee}{\end{equation}}
\newcommand{\diff}[2]{\frac{\partial #1}{\partial #2}}
\newcommand{\con}[2]{\multicolumn{#1}{c}{} &
            \multicolumn{#2}{c}{\vrule width\arrayrulewidth depth0mm
            \hrulefill \vrule width\arrayrulewidth depth0mm}&}
\newcommand{\nl}{\\[-1.6ex]}
\newcommand{\contract}[3]{{
            \renewcommand{\arraystretch}{0.7}
            \hskip-0.4em\begin{array}[t]{*{#1}{c@{\hskip0em}}c}
            #2 \\[-0.7ex] #3 \\ \end{array}\hskip-0.4em}}
\newcommand{\trivcontract}[1]{\contract{4}{#1}{\con{1}{3}}}
\newcommand{\doublecontract}[1]{\contract{8}{#1}{\con{1}{3}\con{1}{3}}}
\newcommand{\crosscontract}[1]{\contract{8}{#1}{\con{1}{5}\nl\con{3}{5}}}
\newcommand{\nestcontract}[1]{\contract{8}{#1}{\con{3}{3}\nl\con{1}{7}}}
\title{RPA-Approach to the Excitations of the Nucleon 
\\ Part I: Theory}
\author{ S. Hardt, J. Geiss, H. Lenske and U. Mosel \\
Institut f\"ur Theoretische Physik \\ Universit\"at Giessen \\
Heinrich-Buff-Ring 16 \\ D-35392 Giessen, Germany}
\maketitle
\begin{abstract}
In this paper we develop a theoretical framework which
allows us to study excitations of the nucleon. Assuming an 
effective two-body interaction as a model for low-energy QCD,
we derive a relativistic TDHF equation for a many-body system
of quarks. 
To render the Dirac-sea contribution to
the mean field finite, we introduce a symmetry conserving
regularization scheme. 
In the small amplitude limit we derive an
RPA equation. 
The structure of the $ph$ interaction and modifications
due to the regularization scheme are discussed.
We give a prescription to obtain a nucleon state with good 
angular momentum ($J$) and isospin ($T$)
quantum numbers on mean-field level.
To study excitations, we develop a tensor-RPA approach, which
is an extension of the conventional RPA techniques to systems
with a nonscalar ground state. 
This allows us to
construct excited states with good $(J/T)$ quantum numbers.
We discuss a method to reduce the overcomplete $ph$-space 
and compute the tensor-RPA interaction matrix elements.
Finally we extend our scheme to include 
$ (\frac{3}{2}^+, \frac{3}{2})$-states. 
\end{abstract}

    
\section{Introduction}
While lattice QCD is on the verge of making definite predictions
for certain quantities like the static $ q \bar{q} $ potential
or the order of the chiral phase transition, hadron spectroscopy
still remains a difficult task due to finite size effects
and uncertainties connected with the valence approximation \cite{Wein94}.
Therefore the need for effective models of low energy QCD persists, especially
in the hadronic sector. Over the years, such approaches have
led to a deeper insight into various aspects of the relation
between the hadronic world and the underlying field theory, which
is formulated with quark and gluon degrees of freedom. \\
We want to focus our attention on models with quark degrees of
freedom only, thus incorporating effects originating from
instantons or monopole condensation at best in an indirect way.
Such fermionic models have been quite successful from a phenomenological
point of view \cite{Mos89}. \\
For the baryonic spectrum the nonrelativistic quark model, based on
the work of Isgur and Karl \cite{IK78}, and related models provide
a satisfactory description of the data. Recently interest in the
nonrelativistic quark model was renewed when it was discovered
that a spin-flavor interaction allows for the correct ordering
of the negative parity states with respect to the positive parity ones
\cite{GR95}. \\
The mesonic part of the spectrum, at least in the pseudoscalar and,
with some restriction, in the vector channel, was increasingly well
understood due to work based on the Nambu-Jona-Lasinio (NJL) model
\cite{NJL61,KLVW90}.
In that regard the role of the pion as the Goldstone boson of the spontaneously
broken chiral symmetry was clarified.
The NJL model possesses solitonic solutions in the $B$$=$$1$ sector which
can be viewed as a mean-field description of the ground-state baryons.
On this basis the phenomenology of ground-state
baryons could be reproduced quite well \cite{ARW95,CGGP95}. \\
The aim of the present work is to develop a general 
framework for a field theoretical quark model 
which describes the nucleon ground state within relativistic
Hartree-Fock theory and
allows to study excited states 
within an RPA approach. A subsequent paper is devoted
to the application of our method. \\  
Wheras in the nonrelativistic quark model
the baryon wavefunctions are built from three-quark configurations
exculsively, we also take into account quark-antiquark admixtures
to the wavefunctions. The quark-antiquark components are included
as a coherent superposition to the excitations of the valence shell. \\
In contrast to the bozonization techniques commonly used
in the NJL model, we derive a Dirac-RPA scheme in close analogy
to the approximation schemes known from nonrelativistic
many-body theory. The essence of such an approach is a 
particle-hole expansion around a mean-field ground state. 
We emphasise that the methods we develop in part I of this
paper are to a large extent model independent. Only in part II
we make a specific choice for the interaction. \\
Several versions of a Dirac-RPA approach have been developed
in the context of nuclear-structure physics \cite{DF90,SRM89}.
In such a relativistic RPA approach it is difficult to
achieve a restoration of symmetries that are broken by
the mean-field configuration. We solve this problem on a very
general basis by introducing a symmmetry-conserving
regularization scheme. \\
The aspect of symmetry is crucial for our approach also
in another regard.
Rotational symmetry of the interaction in coordinate and
isospin space leads to the requirement that the
physical states carry good angular momentum $ (J) $ 
and isospin $ (T) $ quantum numbers.
In most attempts to construct a self-consistent
mean-field state for the nucleon the rotational symmetries are
treated in a rather approximate way. Examples are the
Skyrme model \cite{Sky61,ZB86} and the NJL soliton model 
\cite{ARW95,CGGP95} which are based on a hedgehog symmetry.
In such a description the full symmetry group is truncated
to an invariance under simultaneous rotations in coordinate
and isospin space, leading to conservation of the
grand spin $ {\bf G} = {\bf J} + {\bf T} $. \\
In contrast to that we construct our states as tensors
of the full symmetry group $ SU(2)_J \otimes SU(2)_T $.
On mean-field level this is achieved by a simple projection
technique. On RPA level the situation is more complicated.
The constraints set by the invariance properties of the
underlying quark model force us to introduce an extended
RPA scheme, the so called tensor RPA. The tensor-RPA method
has been developed in the context of nuclear structure physics
to study excitations of nuclei with an odd number of protons
or neutrons \cite{RN75}. We generalize this method and adjust it to
the Dirac-RPA theory for the nucleon. As a result we obtain
a powerful scheme which allows us to compute wavefunctions
and excitation energies for states in any $ (J/T) $-channel. \\
The tensor RPA is also a quite efficient method to build
the correct representations of the symmetric group, since
it is based on second quantization. In the nonrelativistic
quark model a complicated classification scheme has to
be introduced to construct representations of the various
symmetries (see, e.\ g.\ \cite{BIL94}). Compared to that the
construction of states via coupled particle-hole operators
constitutes an alternative and more straightforeward scheme
for building the space of $ (J/T) $-projected baryon wavefunctions. \\ 
Thus the focus of this theoretical part is on two novel
techniques we were forced to introduce in order to maintain the
symmetry structure of a quark model for the nucleon:
on the one hand a special regularization scheme which
guarantees the restoration of broken symmetries on RPA level,
on the other hand a tensor RPA scheme which allows for the
construction of states with good angular momentum and
isospin quantum numbers. \\
The paper is divided into the following sections:
In section 2 we derive an equation of motion for the one-body
density matrix of the system, starting from the
Martin-Schwinger hierarchy for time-ordered Green's functions.
In section 3 we discuss the issue of consistent implementation
of a regularization scheme. 
We focus on the Dirac Hartree-Fock problem for the nucleon
in section 4.
Section 5 is devoted to the discussion
of excited states belonging to multiplets of
the group $SU(2)_{J} \times SU(2)_{T}$. The paper closes with
a discussion and a summary of the results.   
\section{Equations of Motion for a Relativistic System of Fermions}
In the framework of field theoretical quark models
the common way to derive many-body equations of motion 
is via bosonization techniques \cite{E76}. There one
considers an Euclidean path integral and derives the equivalent of
the Hartree-Fock approximation by restricting the functional integration
to the stationary phase configuration. The equivalent of RPA can
be derived by allowing for a time dependence of the mean field. \\
We follow a different path, because we want to stay as close
as possible to the usual formulation of many-body physics 
in terms of
multi-particle multi-hole operators acting on a mean-field ground state. \\
Since we do not perform a Wick rotation to Euclidean space-time
we avoid ambiguities with the treatment of vector
potentials \cite{ARW95}.
We start from a model state given by the
Dirac sea with three valence quarks on top. Dynamic correlations are
treated on the level of small-amplitude oscillations of
the one-body density matrix.
In that spirit we derive many-body equations of motion using
a relativistic version of the Martin-Schwinger hierarchy 
\cite{MS59,BM90}. \\
The class of models we consider is defined by a Lagrangian
of the form
\be
L= \int d1 \bar{\Psi} (1) ( i \gamma^{\mu} \partial_{\mu} -m )
\Psi (1) - \frac{1}{2} \int d1 d2 d1' d2' \langle 1 2 | u | 1' 2' \rangle
\bar{\Psi} (1) \bar{\Psi} (2) \Psi (2') \Psi (1') , 
\ee
where $ 1,2,1',2' $ denote the quantum numbers necessary to
label the single particle states, including the time label.
It is assumed that the quarks interact via an effective
two-body potential $u$.
We define time ordered Green's functions by
\begin{equation}
G(1 \ldots n, 1' \ldots n') = \left( \frac{1}{i} \right)^{n}
\langle T [ \Psi(1) \cdots \Psi(n) \bar{\Psi} (n') \cdots
\bar{\Psi} (1') ] \rangle .
\end{equation}
With that we obtain an equation of motion for the two-point function
\begin{equation}
D(1) G(1,1') = \delta(1-1') -i \int d2 d1'' d2'' \langle 1 2 | u |
1'' 2'' \rangle G(1'' 2'', 1' 2^{+} )
\label{eofm1}
\end{equation}
and the adjoint
\begin{equation}
D^{*} (1') G(1,1') = \delta(1-1') -i \int d2 d1'' d2'' \langle 1'' 2'' | u |
1' 2 \rangle G(1 2^{-} , 1'' 2'').
\label{eofm2}
\end{equation}
The Dirac operator $D$ is given by
\begin{equation}
D= i \gamma^{\mu} \partial_{\mu} -m
\end{equation}
and the notation $D(1) G(1,1')$ denotes the action of $D$ on
the first argument of the Green's function. The superscripts
$+(-)$ indicate that the argument has to be evaluated at
an infinitesimally later(earlier) time compared to the
integration variable. \\
Following closely the derivation of time dependent Hartree-Fock
theory given in \cite{KB62}, we 
approximate the $4$-point function by an antisymmetrized
product of $2$-point functions
\begin{equation}
G(1 2, 1' 2')= G(1, 1') G(2, 2') - G(1, 2') G(2, 1') .
\label{factoriz}
\end{equation}
Furthermore we introduce a one-body density matrix by
\begin{equation}
\rho(1,1') = i G(1,1') \gamma_{0} (1'),
\end{equation}
where $\gamma_0$ is the Dirac matrix times
a unit matrix with respect to space-time, color and flavor indices. \\
If $u$ parametrizes a one-boson exchange interaction
with a local vertex we
encounter a divergence in the equations of motion, because in this
case one of the density operators appearing in the integral
is evaluated at the same points in space-time. Thus, the theory
requires regularization.
In section 3 we will discuss a regularization scheme which
strictly preserves the symmetries of the equation of motion. 
For the time
being we perform all algebraic manipulations as they
can be applied to finite quantities. \\
We introduce new time variables by
\begin{eqnarray}
\tau = t_1 -t_{1'}, & & \; \; t = (t_1 + t_{1'})/2.
\end{eqnarray}
Multiplying Eq.\ (\ref{eofm1}) and (\ref{eofm2}) by $ \gamma_0 $
and taking the difference leads to an equation which only
contains a time-derivative with respect to $t$
\begin{eqnarray}
\lefteqn{ [ i \partial_t 
-( \mbox{\boldmath $ \alpha $ \unboldmath} \! \cdot {\bf p} + \beta m) (1)
+(\mbox{\boldmath $ \alpha $ \unboldmath} \! \cdot {\bf p} +\beta m) (1') ] 
\; \rho(1,1') } \nonumber \\
& & = - \int d2 d1'' d2'' \langle 1 2 | \bar{v} | 1'' 2'' \rangle
\rho (1'',1') \rho(2'',2^{+}) \nonumber \\
& & + \int d2 d1'' d2'' \langle 1 2 | \bar{v} | 1'' 2'' \rangle
\rho (1,1'') \rho(2^{-},2''). \label{gentdhf}
\end{eqnarray}
In this expression
\begin{equation}
v \equiv \gamma_0 \otimes \gamma_0 \; u
\end{equation}
plays the role of a two-body potential as defined in a
hamiltonian formulation. 
The shorthand $ \bar{v} $ indicates antisymmetrization
with respect to particles $1$ and $2$.
The tensor product of the two Dirac matrices defines a
two-body operator. In the above derivation
we have assumed that $v$ is hermitian. \\
If the interaction is local in time, we obtain a closed
equation for the evolution in $t$
\begin{equation}
i \partial_t \rho (1,1',t) = \left[ H_{MF} , \rho \right] (1,1',t),
\label{tdhf}
\end{equation}
where the limit
$ \tau=t_1 - t_{1'} \rightarrow 0^{-} $
was taken.
The mean-field Hamiltonian is defined by
\begin{equation}
H_{MF} (1,1',t) = [ \mbox{\boldmath $ \alpha $ \unboldmath} 
\! \cdot {\bf p} + \beta m] (1,1')
+ \int d2 d2' \langle 1 2 | \bar{v} | 1' 2' \rangle
\rho (2',2,t).
\end{equation}
We thus recover the same structure 
as in the TDHF equation known from nonrelativistic
physics \cite{Neg82}. Although not written in a manifestly covariant manner,
Eq.\ (\ref{tdhf}) is Poincar\'e-invariant if the same is true
for the two-body interaction $v$. This can be traced back
to the approximation introduced in Eq.\ (\ref{factoriz}), which
respects the covariance properties of the $4$-point function. \\
In most cases\footnote{There are, however, a few
exceptional cases where no regularization is needed (see, e.\ g.\
\cite{YOPR84}).}
Eq.\ (\ref{tdhf}) is of little use for practical purposes unless we
specify a regularization scheme.
The two-point function will be infinite due to contributions from
the Dirac sea, so we need to filter the spectral density
of negative energy states to obtain finite results. \\
In the next chapter we will discuss a method that allows us to
regularize the TDHF-equation in a consistent way.
This method will lead to the derivation of a self-consistent
Dirac Hartree-Fock + RPA approach.
\section{Regularization of the TDHF equation}
A covariant theory as the one developed in the previous chapter
confronts us with single particle spectra which are unbound from below.
In that case the standard RPA approach for the excitations of the
system proves to be inappropriate.
Even when one considers a mean field that is made up by positive
energy states only, one has to take special care to develop a
consistent RPA scheme, because the response function picks
up contributions from negative energy states 
\cite{DF90}. \\
We use the term `consistent' 
in the following way:  
Given a Hartree-Fock ground state a method to compute the low-lying
excitations of system is consistent if it respects the symmetries
of the underlying theory.
If a symmetry is broken in the ground state, a mode that corresponds
to a symmetry transformation (e.\ g.\ a translation in space) has
to appear at zero excitation energy. Then the spurious modes
separate from the intrinsic excitations. 
A scheme that does not preserve the symmetries will lead to 
unphysical admixtures of spurious modes to the excited
states. \\
In the nonrelativistic case the consistency problem has been
discussed in much detail \cite{RS80,LM80}. The most
important result is that RPA modes corresponding to
symmetry transformations appear at zero excitation energy.
However, a straightforward extension of the nonrelativistic
RPA scheme does not suffice to describe the linear response
of the Dirac sea. Such a scheme does not allow for the
decoupling of spurious modes in the relativistic case. \\ 
If the mean field is exclusively made up from positive energy states
the simplest way to achieve consistency 
is to shift the
negative energy poles of the Feynman propagator
into the lower half plane \cite{DF90}. 
This corresponds to a drastic change of the structure of
Dirac-hole states from backward to forward motion in time.
Effectively this simulates an unoccupied Dirac sea.
As a consequence the RPA equations include contributions
from configurations where a particle is scattered into a
Dirac-hole state. \\
We follow a different path for the regularization of our
time dependent mean-field theory.
On the one hand we include explicitly the effects of
an occupied Dirac sea. 
On the other hand we avoid the appearance of unphysical
states in the $ph$ basis. On the RPA level our
regularization scheme amounts to using a modified (with respect to
the original theory) interaction. The eigenstates
are defined as a superposition of the physical (positive energy)
$ph$ states and their time-reversed analogues. \\
The basic idea of our regularization scheme is to replace the divergent
mean-field Hamiltonian, which is a functional of the one-body
density, by an effective Hamiltonian of the form
\begin{equation}
H_{MF}' (\rho)= H_{MF} ( R [H_{MF}'] \rho R [H_{MF}'] ).
\label{defhmf}
\end{equation}
Thus only contributions to the mean field
from the regularized density
\be
\rho' = R[ H_{MF}' ] \rho R[ H_{MF}' ]
\ee
are taken into account.
$ R[x] $ is a cut-off function with the property
\begin{equation}
R[x] \rightarrow 0, \; \; (x \rightarrow - \infty)
\end{equation}
for a real argument $x$.
The falloff for large negative arguments should be fast enough
to guarantee that the effective particle number of the ground state 
$ \sum_i R^2 [\epsilon_i ] $ remains finite. The sum is to
be taken over the occupied single-particle states with energies
$ \epsilon_i $. 
We assume that the cut-off function can be expanded into 
a power series, which provides the definition of $R[x]$
for an operator-valued argument. 
Written down in a specific model space Eq. (\ref{defhmf})
is an implicit relation between the matrix elements of
the effective Hamiltonian $ H_{MF}' $, which appear on
both sides of the equation. \\
The modified TDHF equation reads
\begin{equation}
i \partial_t \rho = \left[ H_{MF}' ( \rho ) , \rho \right].
\label{newtdhf}
\end{equation}
The new description allows to
solve for a set of states that simultaneously diagonalize $\rho$ and
$H_{MF}'$, corresponding to the stationary limit of Eq.\ (\ref{newtdhf}).
When we write down $ H_{MF}' $ in a basis of such eigenstates, 
contributions from states arbitrarily deep in the Dirac sea
are excluded by the cut-off function $R$.
However, the
density operator itself will still diverge in a basis of eigenstates
of the position operator, corresponding to a completely filled
Dirac sea. \\
The regularization scheme we propose preserves the symmetries of
the underlying interaction $v$. To show this, consider a
unitary transformation
\begin{equation}
\rho(\theta)= U^{\dagger} \rho(0) U,
\end{equation}
with
\begin{equation}
U=exp( i \theta_a S_a).
\end{equation}
$S_a$ are the generators of any group of transformations that
leave the interaction invariant.
Let us assume that $\rho (0)$ is a solution of Eq.\ (\ref{newtdhf}).
If we can show that $ H_{MF}'$ transforms as
\begin{equation}
H_{MF}' (\theta) = U^{\dagger} H_{MF}'(0) U,
\label{trans}
\end{equation}
we have clearly proven the invariance property of the
regularized equation of motion.
In the above expression the abbrevation
\be
H_{MF}' (\theta) \equiv H_{MF}' ( \rho(\theta) )
\ee
is used. \\
From the defining equation for $H_{MF}'$ we have
\begin{equation}
H_{MF}' (\theta)= H_{MF} ( R [H_{MF}'(\theta)] \rho(\theta) 
R [H_{MF}' (\theta)] ).
\label{defhtheta}
\end{equation}
The proof of the transformation law Eq.\ (\ref{trans}) is based
on the fact that Eq.\ (\ref{defhtheta}) is to be considered as an
implicit {\em definition} of $ H_{MF}' (\theta) $. 
We assume that we are given a solution $ H_{MF}' (0) $
of this equation for $ \theta=0 $.  
To prove the validity of the transformation law, we insert the
right hand side of Eq.\ (\ref{trans}) for $ H_{MF}' (\theta)$
\begin{equation}
U^{\dagger} H_{MF}'(0) U = H_{MF} (R [U^{\dagger} H_{MF}'(0) U]  \;
U^{\dagger} \rho(0) U R [U^{\dagger} H_{MF}' (0) U] ).
\label{def}
\end{equation}
The cut-off function transforms as
\begin{equation}
R[ U^{\dagger} H_{MF}' (0) U ]= U^{\dagger} R[H_{MF}'(0)] U,
\end{equation}
as one finds from a power-series expansion for $R$.
The transformation law for the original mean-field Hamiltonian is
\cite{RS80}
\begin{equation}
H_{MF} (U^{\dagger} \rho(0) U) = U^{\dagger} H_{MF} (\rho(0)) U.
\end{equation}
With that Eq.\ (\ref{def}) becomes
\begin{equation}
U^{\dagger} H_{MF}' (0) U = U^{\dagger} H_{MF} ( R[ H_{MF}' (0) ]
\rho(0) R[ H_{MF}' (0) ] ) U,
\end{equation}
which is the defining equation for the regularized mean-field
Hamiltonian at $ \theta=0 $. Thus the expression given
in Eq.\ (\ref{trans}) is a solution of the defining equation
whenever the same is true for $ H_{MF}' (0) $.
Note that a proof of the transformation law can only be
given in such an indirect manner, since Eq.\ (\ref{defhtheta})
can in general not be solved for $ H_{MF}' (\theta) $. \\ 
Our regularization scheme relies on the fact that the cut-off is
provided by a description which is self-contained, i.\ e.\ with
no reference to an external set of basis states. Had we done an
expansion with respect to any complete set of states,
e.g. momentum eigenstates,
we would have violated the symmetry requirements if we had cut off
the high momentum tails of the density matrix. \\
In fact we have introduced a whole set of regularization schemes,
because so far the form of
the cut-off function $ R $ has not been specified. Clearly
the many-body-theory defined above is not unique and in general
we expect the results to depend on the cut-off function.
Any parametrization which includes the effects of the most
important states close to the fermi surface should, however,
lead to similar results for physical observables.
This can only be checked by applying the relativistic
TDHF theory to a specific model. A similar situation is
encountered in the NJL-soliton model, where different
schemes are used to regularize the diverging contributions from
the Dirac sea. \\    
We now proceed to develop an RPA equation by considering the
small amplitude limit of Eq.\ (\ref{newtdhf}). For this purpose we write
the density operator as
\begin{equation}
\rho= \rho_0 + \delta \rho.
\end{equation}
$ \rho_0 $ is time independent and fulfills
\begin{equation}
\left[ H_{MF}'(\rho_0), \rho_0 \right]=0,
\label{stathf}
\end{equation}
$ \delta \rho $ is a small perturbation.
Inserting this ansatz into the TDHF equation and keeping terms 
linear in $ \delta \rho $ we obtain
\begin{equation}
i \partial_t \delta \rho =
[ \frac{ \partial H_{MF}'}{ \partial \rho} |_{\rho_0}
\delta \rho ,\rho_0 ]  +
\left[ H_{MF}' (\rho_0) , \delta \rho \right]
\label{linearize}
\end{equation}
In the unregularized RPA we would have obtained an expression where
$ H_{MF}'$ is simply replaced by $ H_{MF} $.
The second term of Eq.\ (\ref{linearize}) contains the
information about single-particle energies, since we
can make use of the fact that $ H_{MF}' $ and $ \rho_0 $
are simultaneously diagonal.
The derivative in the first term defines the $ph$ interaction
which enters the RPA equation
\begin{equation}
\langle i j | \bar{v}_{eff} | i' j' \rangle \equiv
\frac{ \partial (H_{MF}')_{i i'}}{ \partial \rho_{j j'}}.
\end{equation}
A similar definition of an effective interaction enters
the Landau-Migdal theory of Fermi liquids \cite{BBN85}. \\
In the standard derivation of the RPA equation
one uses the fact that all particle-particle and hole-hole
matrix elements of $ \delta \rho $ vanish \cite{RS80}. In case that
the mean-field ground state can be written as the vacuum of a set
of single-particle annihilation operators the same statement will
hold for the Dirac RPA. \\
The nucleon, which is a system built from a filled Dirac sea with
three valence quarks on top, requires some special care.
If we model the nucleon by single-particle wavefunctions 
taken from a mean-field calculation, we have to introduce explicit
correlations to construct a $(J$=$1/2$,$T$=$1/2)$ ground state.
In a strict sense this
does not allow us anymore to define the terms `particle' and `hole'.  
However, the old concepts will remain
unaltered for all states outside of the valence shell.
Since such ground-state correlations 
are restricted to the
valence shell, we do not expect to make a big mistake if
for the derivation of the RPA equation we ignore them. \\
A similar situation is encountered in nuclei away from
magic numbers. Already by statistical reasons such systems are
characterized by ground-state occupation probabilites
less than unity in the valence shells. Superimposed
are dynamical correlations due to residual interactions.
In the theory of Fermi-systems the various contributions
to such ground-state correlations are being subsumed into
the quasiparticle concept \cite{Mig68}. This picture takes into
account that the mean-field model states, 
e.\ g.\ from Hartree-Fock theory, are distributed
over a range of many-body eigenstates. \\
We now focus our attention on the effective $ph$ interaction
which originates from the regularization of our theory.
In order to compute the interaction matrix elements we have to examine
the derivative of the regularized mean-field Hamiltonian with
respect to the density. The derivative has to be evaluated at $ \rho_0 $,
the stationary solution of the TDHF equation.
In the following all single particle labels refer to states of
the self-consistent basis, i.\ e.\ the set of states which
simultaneously diagonalize the regularized mean-field Hamiltonian
and the density matrix. As shown in Appendix A, the implicit
equation for the effective interaction is obtained as
\begin{eqnarray}
\lefteqn{ \left. \frac{ \partial}{ \partial \rho_{k k'} } 
H_{MF}' \right|_{\rho_0} =
\diff{ H_{MF}}{ \rho_{k k'}' } R[\epsilon_k] R[\epsilon_{k'} ] }
\nonumber \\
& & + \left. \diff{H_{MF}}{\rho_{l l'}'} \right|_{\rho_0}
\left. \diff{ \left( H_{MF}' \right)_{l l'} }{ \rho_{k k'}} 
\right|_{\rho_0} \frac{R[\epsilon_{l'} ]- R[\epsilon_{l}] }{ \epsilon_{l'}
- \epsilon_l } \left( \theta (k_F -l) R[ \epsilon_l]
+ \theta (k_F -l') R[ \epsilon_{l'}] \right).
\label{effint} 
\end{eqnarray}
As an example which reveals the structure
of the above equation let us consider a cut-off function of the
form
\be
R_l [\epsilon_i] = e^{-e^{-l ( \epsilon_i
+ \Lambda )}},
\label{simple}
\ee
where $l$ has dimension $ [ MeV^{-1} ]$.
Clearly $ R_l $ has a power series representation and
\be
\lim_{l \rightarrow \infty} R_l [\epsilon_i] = \theta(\epsilon_i+ \Lambda). 
\ee
In the following let us restrict ourselves to very large $l$,
where the step function is reproduced with high accuracy.
We want to emphasize that the above choice for the cut-off function
is made in order to provide an interpretation of Eq. (\ref{effint})
in terms of $ph$ states and to compare our approach to
the previously mentioned Dirac-RPA method of Dawson and Furnstahl
\cite{DF90}. Their approach allows for particle states in the Dirac sea.
At the moment we do not touch the question 
if the special choice for the cut-off function in Eq.\ (\ref{simple}) 
is suited for practical purposes, because our aim is to illuminate
the structure of the equations rather than going into details
of a numerical solution. In fact, as has been shown in Ref.\
\cite{BDMG90}, a sharp cut-off might lead to problems when
attempting to solve self-consistent mean-field equations
by numerical methods. \\
We write the matrix elements of the effective interaction as
\be
V_{k k' , l l'} \equiv \diff{(H_{MF}')_{k k'} }{ \rho_{l l'} }.
\ee
When two-body states $ \alpha=(k k'), \; \beta=(l l') $ 
are introduced 
the defining equation for the effective interaction can be
written as a matrix equation of the form
\be
V_{\alpha \beta} = M_{\alpha \mu} 
V_{\mu \beta} + N_{\alpha \beta},
\label{shorthand}
\ee
where the matrices $M$ and $N$ can be read off from Eq.\ (\ref{effint}). 
We introduce model spaces $ {\cal E} $ and $ {\cal D} $ of two-body
states defined by
\begin{itemize}
\item[ $ {\cal E}$: ] $ph$ states $(k k')$ with 
$ \left\{ \begin{array}{c} \epsilon_k  >  \epsilon_F \\
\epsilon_{k'}  <  \epsilon_F \end{array} \right\} $ or 
$ ( k \leftrightarrow k' )$ 
\item[ $ {\cal D}$: ] states $(k k')$ with $ \left\{ \begin{array}{c}
- \Lambda < \epsilon_k < \epsilon_F \\
\epsilon_{k'} < -\Lambda \end{array} \right\} $ or $ ( k \leftrightarrow
k' ) $.
\end{itemize}
$ \epsilon_F $ denotes the Fermi energy.
Clearly the spaces $ {\cal E} $ and $ {\cal D} $ are orthogonal. \\
Let $ P $ and $ Q $ be the projectors onto $ {\cal E} $
and $ {\cal D} $.  
For an arbitrary operator $A$ we introduce the notation
$ A_{PP} \equiv P A P, \; A_{QQ} \equiv Q A Q $,
similarly for other combinations of $P$ and $ Q$. 
Note that $P$ and $Q$ are operator-valued functionals of the
self-consistent density $\rho_0 $. \\
The structure of $M$ in Eq.\ (\ref{shorthand}) is such
that the sum over $ \mu $ is restricted to states in $ {\cal D} $.
This fact is due to the special choice for the cut-off function
Eq.\ (\ref{simple}).
With that we obtain the two relations
\begin{eqnarray}
V_{PP} & = & M_{PQ} V_{QP} + N_{PP} \nonumber \\
V_{QP} & = & M_{QQ} V_{QP} + N_{QP}. 
\end{eqnarray}
When we solve the second equation for $ V_{QP} $ and insert
the solution into the first equation, we find
\be
V_{PP} = M_{PQ} (1- M_{QQ})^{-1} N_{QP} + N_{PP}.
\label{symbol}
\ee
Similar structures involving projection operators are encountered
in the theory of valence forces in nuclei \cite{DF74}.
There one considers the modification of the interaction
due to the truncation of the Hilbert space to a
smaller model space. \\ 
For the TDHF equation we exclusively need $ph$ matrix elements
of $V$, i.\ e.\ matrix elements in $ {\cal E} $.
The model space $ {\cal D} $ contains the de-excitation states
of Dawson and Furnstahl \cite{DF90}, which can be viewed as
negative energy $ph$ states with respect to a `Fermi surface'
at cut-off energy. In contrast to their prescription the
de-excitation states only enter as intermediate states in our
RPA scheme and do not contribute to the collective $ph$ amplitudes
describing the excited states of the system. \\
When the RPA eigenvalue equation is solved with the effective
$ph$ interaction defined by Eq.\ (\ref{symbol}), the
spurious modes decouple from the excitation spectrum.
\section{Mean-Field Theory of the Nucleon}
For the ground state of the nucleon we have to solve the stationary
limit of the regularized TDHF equation, which amounts to finding
the eigenvalues of the regularized mean-field Hamiltonian.
The fact that $H_{MF} $ is replaced by $ H_{MF}' $ formally
leads to a highly nonlinear problem. The eigenvalues are, however,
determined iteratively. \\
In order to exemplify the procedure without the quite complicated
valence-shell contributions to the mean field, let us 
consider the analogous problem for the vacuum.
Initially we choose a set of states which should not be too far
away from the self-consistent set in order to guarantee convergence.
We then compute the regularized
mean-field Hamiltonian assuming that the states we have chosen
are eigenstates.
The Hartree-Fock potential of the vacuum acquires the form
\be
V_{MF}= \sum_{pqi} \langle p i | \bar{v} | q i \rangle
a^{\dagger}_p a_q R^2 [\epsilon_i] \theta (-\epsilon_i) ,
\ee   
where in the argument of the regularization function $ H_{MF}' $
has been replaced by an approximation for its eigenvalues.
After that we solve the
eigenvalue problem for the new $ H_{MF}' $ and repeat the procedure
until convergence is achieved. By construction a configuration
obtained in such a way represents a solution of Eq.\ (\ref{stathf}). \\
A well known model for vacuum structure
and spontaneous chiral symmetry breaking is the Nambu-Jona-Lasinio(NJL) model
(for a review, see e.\ g.\ \cite{Kle92}). Due to the simple momentum structure
of the NJL model it is easy to derive a nonlinear equation
which determines the
chiral condensate $ \langle \bar{\Psi} \Psi \rangle $.
With the assumption of a
translationally invariant vacuum state this so called gap equation
can also be derived within our Dirac-Hartree-Fock scheme
applied to the NJL model. One finds
\be
\langle \bar{\Psi} \Psi \rangle \propto \int_{m}^{\infty} d E
\sqrt{ E^2 - m^2 } m R^2[E],
\label{gap}
\ee
where $ m $ is the constituent quark mass which includes contributions
from the scalar part of the mean-field potential.
Eq.\ (\ref{gap}) is identical to the gap equation with an
$ O(3) $-invariant cut-off discussed in \cite{Kle92}
when a step-function parametrization is chosen for $ R^2 [E] $. \\
From this discussion it should become clear that the Dirac-Hartree-Fock
method set forth in the previous chapter contains previously developed
methods for the determination of the mean-field structure of 
fermionic systems. The special form of the regularization scheme
based on an implicit definition of the mean-field Hamiltonian
mainly has an effect on the RPA sector of the theory where it leads to
a modification of the $ph$ interaction. \\
Compared to the vacuum sector additional complications are introduced
when one includes valence quarks and studies a nucleonic system. 
As mentioned in the introduction, our aim is to construct states that
carry good $J/T $ quantum numbers. However, the $ J/T $ symmetries
are not automatically self-consistent symmetries of the nucleon
mean-field Hamiltonian. This means that the mean-field of
a system described by a many-body state with good $J/T$
quantum numbers is no longer invariant under rotations in
coordinate and isospin space.  
Subsequently it will be only be possible
to fulfill the symmetry requirements by acting with a projection
operator on the eigenstates of this symmetry-broken
Hamiltonian. \\
A possible approach would be to define the nucleon as
a state with grand spin $ G=0$,
where the grand spin is defined as
\be
{\bf G} = {\bf J} + {\bf T}.
\ee
This so called `hedgehog' ansatz is used in the chiral soliton models,
because the grand spin symmetry is a self-consistent symmetry
of the mean-field Hamiltonian.
After solving the mean-field equation states with good $J/T $ quantum
numbers are projected out, usually by a cranking procedure.
In the language of nonrelativistic nuclear physics this
approach would be called a `variation before projection' method 
\cite{RS80}. \\
A superior but technically more involved approach is provided
by the `variation after projection' techniques.
There the mean-field approximation is applied to a projection
of the full hamiltonian onto a subspace of $J/T$ eigenstates 
(see \cite{SGF184,SGF284} and references therein).
The projected ground-state wave function can in general
only be written as a superposition of several slater determinants. \\
In order to keep the theoretical discussion more transparent
and numerical investigations feasible we restrict ourselves to
a somewhat simplified `variation after projection' technique
in the following.
We assume that the valence shell is a
$ \kappa =1 $ or $ \kappa=-1 $ state, thus carrying an
angular momentum of $ J=1/2 $. Although the notation we use
suggests a valence level of positive energy, the positivity
of the valence energy is not mandatory for our discussion.
The vacuum $ |0 \rangle $
which we refer to as `Dirac sea' is defined by the 
single-particle states which are lower in energy than
the valence level. \\
A wavefunction with the quantum numbers of the nucleon
($ (J^p,T)= (\frac{1}{2}^+,\frac{1}{2}) $)
is given by \cite{Mos89}
\be
|\Gamma_N , \Gamma_z \rangle = \frac{1}{ \sqrt{18}} 
\sum_{\Gamma_{z1}, \Gamma_{z2}, \Gamma_{z3}}
T_{\Gamma_{z1} \Gamma_{z2} \Gamma_{z3}}^{( \Gamma_z)}
a_{\Gamma_{z1}}^{1 \dagger} a_{\Gamma_{z2}}^{2 \dagger} 
a_{\Gamma_{z3}}^{3 \dagger} | 0 \rangle .
\label{nucl}
\ee
The quark states are labelled by their color quantum number (superscripts)
and angular-momentum/isospin quantum number (subscripts).
The $ (J,T) $ quantum numbers of the nucleon are denoted with
$ \Gamma_N \equiv (1/2,1/2) $, the projection quantum numbers
with $ \Gamma_z $.
In the following we take the convention that doublets of $(J/T)$
quantum numbers are subsumed in one single symbol
\be
\Gamma_i \equiv (J_i, T_i).
\ee
Projection quantum numbers are abbreviated in a similar way
\be
\Gamma_{zi} \equiv ((J_z)_i ,(T_z)_i) .
\ee
Unless stated otherwise, we define for any function
of the $(J/T)$ doublets
\be
f(\Gamma_1, \Gamma_2, \ldots, \Gamma_n) \equiv
f(J_1, J_2, \ldots, J_n) f(T_1, T_2, \ldots, T_n ) .
\label{split}
\ee
State vectors and tensor operators which are labelled
by $ \Gamma $-symbols are an exception to this convention.
To express the nucleon wavefunction in terms of creation operators
we have introduced coupling coefficients
$ T_{\Gamma_{z1} \Gamma_{z2} \Gamma_{z3}}^{( \Gamma_z)} 
$ of the group $ SU(2)_{J} \otimes
SU(2)_{T} $ \cite{Mos89}.
The $T$-coefficients {\em do not} split up according to Eq.\ 
(\ref{split}). \\
For the further discussion it is appropriate to introduce a creation
operator for a 3-quark state with nucleonic quantum numbers
\be
A^{\dagger}_N (\Gamma_z ) \equiv 
\frac{1}{ \sqrt{18}} 
\sum_{\Gamma_{z1}, \Gamma_{z2}, \Gamma_{z3}}
T_{\Gamma_{z1} \Gamma_{z2} \Gamma_{z3}}^{( \Gamma_z)}
a_{\Gamma_{z1}}^{1 \dagger} a_{\Gamma_{z2}}^{2 \dagger} 
a_{\Gamma_{z3}}^{3 \dagger}
\ee
By definition this operator fulfills
\begin{eqnarray}
A^{\dagger}_N (\Gamma_z) |0 \rangle & = & | \Gamma_N, \Gamma_z \rangle
\nonumber \\
A_N (\Gamma_z) | 0 \rangle & = & 0 .
\label{aadagger}
\end{eqnarray}
The diagonal matrix elements of the density operator defined with
respect to a nucleon state with projection quantum numbers
$ \Gamma_z $ then become
\begin{eqnarray}
\rho^{(\Gamma_z)} (m) & = & \langle 0 | A_N ( \Gamma_z ) \Psi^{\dagger}
(m) \Psi (m) A^{\dagger}_N ( \Gamma_z ) | 0 \rangle \nonumber \\
& = & \langle 0 | \Psi^{\dagger} (m) \Psi (m) | 0 \rangle
+ \langle 0 | [ [ A_N ( \Gamma_z) , \Psi^{\dagger} (m) \Psi (m) ],
A^{\dagger}_N ( \Gamma_z ) ] | 0 \rangle .
\label{val+sea}
\end{eqnarray}
From this expression it becomes obvious that the density operator
contains contributions from the Dirac sea and from the valence
quark configuration, respectively. It is easy to verify Eq.\ 
(\ref{val+sea}) by writing out the double commutator and
making use of Eq.\ (\ref{aadagger}).
The different contributions are evaluated as \cite{Har92}
\be
\rho^{( \Gamma_z)} (m) =  g^{( \Gamma_z) } (m) / 18 + \theta(-m),
\label{dens}
\ee
where the symbolic $\theta$ function is defined as
\be
\theta(-m) = \left\{ \begin{array}{lcl} 1 & : & m \; \epsilon \; 
$Dirac sea$ \\ 0 & : & $otherwise$ \end{array} \right. .
\ee
The valence-occupation factors $g$ are color independent and given by
\be
g^{( \Gamma_z)}(m) = \left\{ \begin{array}{lcl} 
\sum_{\Gamma_{z1}, \Gamma_{z2}} 
(T_{ \Gamma_{zm} \Gamma_{z1} \Gamma_{z2}}^{( \Gamma_z)})^{2} & : 
& m \; \epsilon \; $valence shell$ \\
0 & : & $otherwise$ \end{array} \right. .
\label{occprob}
\ee
From these relations it is apparent that the valence part of
the full wavefunctions leads to a dependence of $ \rho^{( \Gamma_z)} (m)$
on the $ J/T $ projection quantum numbers of $m$.
In a basis of eigenstates of $ H_{MF}' $ the mean-field potential
for a nucleon with projection quantum numbers $ \Gamma_z $ is defined as
\be
V_{MF} = \sum_{p q i} \langle p i | \bar{v} | q i \rangle
a_{p}^{\dagger} a_q R^2[ \epsilon_i ] \rho^{( \Gamma_z)} (i).
\label{vmf}
\ee
Due to the noninvariant structure of the density matrix
the mean-field potential is no longer invariant under rotations
in coordinate and isospin space. Such complications are well
known from nuclear structure physics and have been treated on
mean-field level allowing for angular momentum and isospin mixing
in the wavefunctions \cite{Pas76}. \\
In the present context the approximation scheme we use to
describe the excited states requires a single particle
basis of $J/T$ eigenstates.
In order to enforce that the mean-field potential is invariant with respect
to rotations in coordinate and isospin space we 
use occupation probabilities averaged over projection
quantum numbers
\be
\rho^{( \Gamma_z)} \rightarrow \bar{\rho} \equiv \frac{1}{4} 
\sum_{ \Gamma_z } \rho^{( \Gamma_z)}.
\label{avrho}
\ee
This approximation neglects the nonscalar parts of the
density matrix.
As a result, the averaged density becomes independent
of the projection quantum numbers and is given by
\be
\bar{\rho} (m) = \frac{1}{4}
\label{valocc}
\ee
for any state $m $ of the valence shell. \\
Comparing this prescription to the usual hedgehog ansatz we
find that $ \bar{\rho} $ contains contributions from
configurations with $ G \not= 0 $. Due to the unconstraint
summation over angular momentum and isospin projection
quantum numbers $ G_z $ can take the values $ 0, \pm 1 $. \\
Another difference with respect to the hedgehog ansatz is related to
the Lorentz-structure of the mean field. Due to the averaging
procedure only the Lorentz-scalar and the time-like component
of the Lorentz-vector part contributes. Pionic contributions
to the stationary mean field which are a major building block of the
hedgehog soliton are thus excluded. Note, however, that before
the averaging the expectation value $ \langle \Gamma_N , \Gamma_z |
\bar{\Psi} \gamma_5 \mbox{\boldmath$\tau$\unboldmath} \Psi | \Gamma_N,
\Gamma_z \rangle $ can take a nonzero value. For the same reason
the pion field will contribute in the more
elaborate projection scheme of Refs.\ \cite{SGF184,SGF284}. 
With such a description the tensor-RPA
matrixelements to be discussed in the following section will exhibit a
quite complicated structure. Thus in the present work we restrict ourselves
to the averaging prescription of Eq.\ (\ref{avrho}). \\ 
Within this simplified scheme the pionic contributions are recovered
on the level of ground-state correlations. 
Such correlations are inherent in an RPA description,
where the information on improvements of the ground state
beyond the static mean-field level
enters via the $ Y $-amplitudes. In \cite{GLM96}
a simple RPA scheme for the nucleon was proposed where
the tensor structure of the excited states is ignored.
In such a scheme the single particle occupation probabilities
for the RPA ground state are given by
\begin{eqnarray}
n_h & = & 1 - (2 \Gamma_h +1)^{-1} \; \frac{1}{2} \sum_{\Gamma, \Gamma_z,
p, \nu} (2 \Gamma +1) | Y^{\nu}_{ph} (\Gamma, \Gamma_z) 
|^2 ( \bar{\rho} (h) - \bar{\rho} (p) )^2 \nonumber \\
n_p & = & (2 \Gamma_h +1)^{-1} \; \frac{1}{2} \sum_{\Gamma, \Gamma_z,
h, \nu} (2 \Gamma +1) | Y^{\nu}_{ph} (\Gamma, \Gamma_z)
|^2 ( \bar{\rho} (h) - \bar{\rho} (p) )^2 ,
\end{eqnarray} 
where $n_h$ and $ n_p $ are the occupation probabilities for
hole and particle states, respectively. The backward amplitudes
$ Y $ are defined in the usual way (see, e.\ g.\ \cite{RS80}). 
The depletion of hole states and occupation of particle states
is due to $2p$-$2h$-admixtures to the ground state. 
These admixtures contain contributions from $ph$
excitations of the vacuum, where the pionic modes are
expected to be most important. In \cite{GLM96} it has
been shown that the vacuum contributions are in fact
sizeable, leading to a depletion of the Dirac sea of
about $ 10 \% $. \\
With such an improved description of the ground state not
only the coupling to pionic modes is recovered, but it
is also possible to study the spin structure of the nucleon.
In the nucleon wavefunction of Eq.\ (\ref{nucl}) the spin 
is carried by the valence quarks exclusively. When
ground-state correlations are taken into account, a part
of the angular momentum is transferred to $ph$ admixtures. \\ 
The ground state density of Eq.\ (\ref{val+sea})
or Eq.\ (\ref{avrho}) does not 
allow an unambiguous assignment of particles and holes,
since the valence shell is only partially filled and the
vacancy can be populated in excitations.
The structure of the $ph$ operators to be included in the
RPA basis will be discussed in the next section. \\
Applying Eq.\ (\ref{avrho}) to a valence quark configuration
coupled to $ \Gamma_{\Delta} \equiv
(3/2,3/2) $ the same averaged occupation probabilities
as in Eq.\ (\ref{valocc}) are found. Thus, the $ (1/2^+,1/2) $
and the $ (3/2^+,3/2) $ valence quark configuration are indistinguishable
on the level of the averaged density matrix. However, as discussed
in the next section, this degeneracy will be lifted from
configuration mixing from residual interactions.
When we take
the expectation value of the full Hamiltonian with an explicit
representation of the ground state like Eq.\ (\ref{nucl}) we
in general find different values for nucleon and $ \Delta $.
This means that the expectation value of the residual interaction
is nonzero for both states due to ground-state correlations.
For the derivation of the RPA equation in the previous chapter
we have assumed that these correlations are neglible, i.\ e.\
we have derived the equation in the same manner as for a closed-shell
system. A measure for the importance of correlations is the
difference $ | \langle \Gamma_N | H | \Gamma_N \rangle -
\langle \Gamma_{\Delta} | H | \Gamma_{\Delta} \rangle | $. If
this difference is not small compared to the lowest single
particle excitation energies a RPA scheme assuming an uncorrelated
ground state must be taken with precaution. \\ 
The important point of our description  
is the fact that we can extract a single-particle
basis with good $J/T$ quantum numbers from the mean-field Hamiltonian
defined with the averaged density.
This will be a crucial prerequisite for the tensor-RPA method
discussed in the next section.
\section{Tensor RPA}
The tensor-RPA approach is a special case of the tensor equations of motion
method developed more than 20 years ago
by Rowe et al.\ \cite{RN75}, which, however,
never seems to have been used in realistic calculations.
It allows to compute excited states of a system with a nonscalar (with
respect to a $SU(2)$-symmetry) ground state.
Usually RPA calculations are restricted to closed-shell systems with
a spin/isospin zero ground state. In this case the excited states
carry the quantum numbers of the $ph$ configuration
admixed to the ground-state wavefunction. \\
For a nonscalar system the situation is more complicated.
We assume that the Hamiltonian which governs the dynamics
is invariant under spatial rotations ($SO(3)$) and isorotations
($SU(2)$). Since $SO(3)$ and $SU(2)$ are isomorphic, the
invariance group is $SU(2)_J \otimes SU(2)_T$. 
The eigenstates of the Hamiltonian should transform as
tensors with respect to this symmetry.
An excited state of given $ J/T $ quantum numbers can in general be
defined by coupling $ph$ pairs of different tensor ranks
on the ground state, according to the rules for outer
products of $SU(2) $ representations. \\
For the nucleon we have to set up such an extended RPA formalism
in order to compute excitations. For the time being we want
to stay in the $SU(2)$ sector of the flavor group.
This seems to be a reasonable approximation, because hints
for strangeness in the excitation spectrum of the nucleon
are restricted to the $N(1535)$ resonance, which decays to a sizeable
amount into a nucleon and an $ \eta$ meson.
A $SU(3)$ tensor RPA could be formulated as a straightforward
extension of the method discussed here, but in order to
apply it to baryon phenomenology one would have to
think of a mechanism which splits the multiplets into
several $SU(2)$ representations and which can be traced
back to a symmetry breaking mass matrix \cite{GR95}. \\ 
The formal derivation of the tensor-RPA equation
for the nucleon is independent of a specific representation
of the ground state. The general formalism is therefore
applicable also in the context of a more sophisticated
description than the one used in the previous chapter. 
Details of the method as the number of linearly independent
basis states or the structure of the interaction matrix
elements, however, depend on the ground-state structure.
\subsection{The Tensor-RPA Equation}
We assume that an excited state of the nucleon is defined
by the action of an excitation operator on the ground state
\be
| x \Gamma_{\Delta} \rangle \rangle = ( Q_{ \Delta }^{\dagger} \times
| \Gamma_N \rangle \rangle )^{\Gamma_{\Delta}}.
\label{exstate}
\ee
$ \Gamma_{\Delta}( \Gamma_N )$ denotes the 
$SU(2)_J \otimes SU(2)_T$ tensor rank
of the excited state (ground state) and $x$ includes all
additional quantum numbers.
Since the adjoint of a tensor state is found to transform
as a tensor only when supplied with appropriate
phase factors \cite{RN75}, we use the definition
\begin{eqnarray}
& & | \Gamma, \Gamma_z \rangle \rangle \equiv 
| \Gamma, \Gamma_z \rangle \nonumber \\
& & \langle \langle \Gamma, \Gamma_z | \equiv (-1)^{\Gamma_z} \langle
\Gamma, -\Gamma_z |,
\end{eqnarray}
where $\Gamma_z$ is the projection quantum number.
A similar definition is used for operators
\begin{eqnarray}
& & O^{\dagger} ( \bar{\Gamma}, \Gamma_z) \equiv O^{\dagger} 
(\Gamma, \Gamma_z) \nonumber \\
& & O (\bar{\Gamma}, \Gamma_z) \equiv (-1)^{\Gamma + \Gamma_z}
O ( \Gamma, - \Gamma_z) .
\end{eqnarray}
The notation for the tensor coupling we use in Eq.\ (\ref{exstate}),
applied to arbitrary tensors $R$ and $S$, stands for
\be
\left( R^{\Gamma_1} \times S^{\Gamma_2} \right)^{\Gamma}_{\Gamma_z} \equiv
\sum_{\Gamma_{z1} \Gamma_{z2}} \langle 
\Gamma_1 \Gamma_2 \Gamma_{z1} \Gamma_{z2} | \Gamma \Gamma_z
\rangle R^{\Gamma_1}_{\Gamma_{z1}} S^{\Gamma_2}_{\Gamma_z2} ,
\ee
with Clebsch-Gordan coefficients that split up according to
Eq.\ (\ref{split}).
The excitation operator $Q_{ \Delta}^{\dagger}$ contains 
a sum over several tensor ranks $ \Gamma_i $ 
\be
Q_{ \Delta}^{\dagger} = \sum_i O_{x \Gamma_i }^{\dagger} ,
\ee
according to angular-momentum/isospin
selection rules.
We demand
\be
Q_{ \Delta} | \Gamma_N \rangle \rangle = 0
\ee
to ensure that the excited states are orthogonal to the ground state.
Furthermore we assume that the excitation operators can be written
as a superposition of certain operators $ \eta^{\dagger} $ yet
to be specified
\be
O_{x \Gamma_i }^{\dagger} = \sum_{\beta} Z_{\beta i} (x)
\eta_{\beta}^{\dagger} (\Gamma_i ).
\ee
Based on these assumptions a general equation of motion for
the excited states of the nucleon can be derived,
which in structure resembles the results of Ref.\ \cite{RN75}
\be
M Z(x) = \omega_{ \Delta} N Z(x),
\ee
where $ \omega_{ \Delta}$ are the excitation energies.
The hermitian matrices $M$ and $N$ are defined as
\begin{eqnarray}
M_{\alpha i, \beta j} & = & \sum_{\Gamma} \hat{\Gamma} 
(-1)^{\Gamma_N - \Gamma_{\Delta}
- \Gamma - \Gamma_i } \langle \Gamma_N || 
[ \eta_{\alpha} ( \bar{\Gamma}_i ),
H , \eta_{\beta}^{\dagger} ( \Gamma_j ) ]^{\Gamma} || \Gamma_N \rangle 
\nonumber \\
& & \cdot W( \Gamma_i \Gamma_j \Gamma_N \Gamma_N, \Gamma \Gamma_{\Delta} ) 
\nonumber \\
N_{\alpha i, \beta j} & = & \sum_{\Gamma} \hat{\Gamma} 
(-1)^{\Gamma_N - \Gamma_{\Delta} - \Gamma - \Gamma_i}
\langle \Gamma_N || [ \eta_{\alpha} ( \bar{\Gamma}_i ),
\eta_{\beta}^{\dagger} ( \Gamma_j ) ]^{\Gamma} || \Gamma_N \rangle \nonumber
\\
& & \cdot W( \Gamma_i \Gamma_j \Gamma_N \Gamma_N, \Gamma \Gamma_{\Delta} ) .
\label{teom}
\end{eqnarray}
In this expression $H$ is the full Hamiltonian of the system and $W$
are the $SU(2)_J \otimes SU(2)_T $ Wigner 6j-symbols \cite{Ros57}.
We have introduced a frequently needed 
symbol for the dimension of multiplets by
\be
\hat{\Gamma} \equiv (2 \Gamma +1 )^{\frac{1}{2}}.
\ee
A coupled commutator of two operators $P$ and $Q$ is defined as
\be
\left[ P^{\Gamma_1}, Q^{\Gamma_2} \right]^{\Gamma}_{\Gamma_z} \equiv
\sum_{\Gamma_{z1} \Gamma_{z2}} \langle \Gamma_1 
\Gamma_2 \Gamma_{z1} \Gamma_{z2} | \Gamma \Gamma_z
\rangle \left[ P^{\Gamma_1}_{\Gamma_{z1}}, 
Q^{\Gamma_2}_{\Gamma_{z2}} \right] .
\label{couco}
\ee
A symmetrized
double commutator will be denoted by
\be
2 \left[ A, B, C \right] \equiv \left[ A, \left[ B, C \right] \right]
+ \left[ \left[ A, B \right], C \right].
\ee
When one of the three operators is a scalar, as in Eq.\ (\ref{teom}),
we use the notation of Eq.\ (\ref{couco}) for the coupling
of the remaining two operators. \\ 
Depending on the choice for the operators $\eta$ one can
derive several approximation schemes from this equation, e.\ g.\
Tamm Dancoff, RPA or quasiparticle RPA.
The states are normalized according to
\be
Z^{\dagger} (y) N Z (x) = \delta_{x y},
\ee
with a metric tensor $N$ which will in general be different from
the identity matrix.
For deriving an RPA equation, we make the specific choice
\be
O_{x \Gamma_k }^{\dagger} = \sum_{m,i} \left[ X_{(mi)k} (x) A_{mi}^{\dagger}
(\Gamma_k ) - Y_{(mi)k} (x) A_{mi} (\bar{\Gamma}_k ) \right] .
\label{rpaop}
\ee
An operator $ A_{mi}^{\dagger} (\Gamma_k) $ creates a $ph$
pair $ (mi)$ with $ J/T$ quantum numbers $ \Gamma_k $ and
is defined as
\be
A_{mi}^{\dagger} (\Gamma_k \Gamma_{zk}) =
\sum_{ \Gamma_{z m} \Gamma_{z i}}
\langle \Gamma_m \Gamma_i \Gamma_{z m} \Gamma_{z i} |
\Gamma_k \Gamma_{zk} \rangle a_{\Gamma_m \Gamma_{zm} }^{\dagger}
a_{ \bar{\Gamma}_i \Gamma_{z i} }.
\ee
We adopt the convention that particle states are denoted by $m,n$
and hole states by $i,j$. 
A state above the valence shell is always a particle, 
a state in the Dirac sea always a hole.
Within the valence shell there can be both types of states. 
This opens up the possibility of having 
$ph$ states where both particle and hole are in the valence shell.
Such states would induce a recoupling of the ground state
and can be used to express the wavefunction of the $ \Delta (1232)$,
which is degenerate with the nucleon on mean-field level.
However, introducing such recoupling states into the set of
excitation operators requires special care.
This issue will be discussed later. \\
With Eq. (\ref{rpaop}) the tensor equation of motion takes the form
\be
\left( \begin{array}{cc}   M^{(1)} & -M^{(2)}  \\
-M^{(3)} & M^{(4)} \end{array} \right) \left( \begin{array}{c}
X \\ Y \end{array} \right) = \omega_{\Delta}
\left( \begin{array}{cc}  N^{(1)} & -N^{(2)} \\
-N^{(3)} & N^{(4)} \end{array} \right) \left( \begin{array}{c}
X \\ Y \end{array} \right) .
\label{tensorrpa}
\ee
With the definition
\be
C_{k,l}^{\Gamma} \equiv (-1)^{ \Gamma_N - \Gamma_{\Delta} 
-\Gamma -\Gamma_k }
\hat{\Gamma} \: W( \Gamma_k \Gamma_l \Gamma_N \Gamma_N, 
\Gamma \Gamma_{\Delta} )
\ee
the submatrices are given by
\begin{eqnarray}
M_{(mi)k,(nj)l}^{(1)} & = & \sum_{ \Gamma} C_{k,l}^{\Gamma}
\langle \Gamma_N || [ A_{mi} ( \bar{\Gamma}_k ) , H , A_{nj}^{\dagger}
(\Gamma_l ) ]^{\Gamma} || \Gamma_N \rangle \nonumber \\
M_{(mi)k,(nj)l}^{(2)} & = & \sum_{ \Gamma} C_{k,l}^{\Gamma}
\langle \Gamma_N || [ A_{mi} ( \bar{\Gamma}_k ) , H , A_{nj}
(\bar{\Gamma}_l ) ]^{\Gamma} || \Gamma_N \rangle \nonumber \\
M_{(mi)k,(nj)l}^{(3)} & = & \sum_{ \Gamma} C_{k,l}^{\Gamma}
\langle \Gamma_N || [ A_{mi}^{\dagger} ( \Gamma_k ) , H , A_{nj}^{\dagger}
(\Gamma_l ) ]^{\Gamma} || \Gamma_N \rangle \nonumber \\
M_{(mi)k,(nj)l}^{(4)} & = & \sum_{ \Gamma} C_{k,l}^{\Gamma}
\langle \Gamma_N || [ A_{mi}^{\dagger} ( \Gamma_k ) , H , A_{nj}
(\bar{\Gamma}_l ) ]^{\Gamma} || \Gamma_N \rangle .
\label{rpamatr}
\end{eqnarray}
We obtain the matrices $N$ when we replace the double commutators
by ordinary commutators and leave away the Hamiltonian $H$.
As a consequence we have
\be
N^{(2)}=N^{(3)}=0 .
\ee
It is understood that we use the nucleon ground state as defined
in Eq.\ (\ref{nucl}) in these expressions. Note, however,
that with a suitably generalized definition of particle
and hole states the description remains valid for
any ground state with the correct tensor structure. \\
So far we have suppressed the color degree of freedom.
According to the confinement property of QCD we demand that
the excited states and the ground state are color singlets.
Then the excitation operator is restricted to color singlet
$ph$ pairs.
We can easily incorporate the color degree of freedom
in Eq.\ (\ref{rpamatr}) by making the replacement
\be
a_{i}^{\dagger} a_m \rightarrow \frac{1}{\sqrt{3}} \sum_{c}
a_{ic}^{\dagger} a_{mc}.
\ee
A similar expression holds for the adjoint $ph$ operators.
We will display the color quantum numbers only when they are
explicitly needed. \\
In case of a scalar ground state with respect to the group
$ SU(2)_J \otimes SU(2)_T $ the tensor-RPA equation (\ref{tensorrpa})
can be shown to reduce to an ordinary RPA equation \cite{Har92}. \\
Given a real interaction, we find that $M^{(1)} $ is equal to $ M^{(4)} $
and that $ M^{(2)} $ is symmetric in case of a time-reversal invariant
(in terms of $SU(2)_J$) system, i.\ e.\ a scalar ground state.
For a nonscalar ground state these relations are
found not to be fulfilled anymore.
However, restricting
the summation in Eq.\ (\ref{rpamatr}) to $ \Gamma=(0,0) $ we 
recover symmetric expressions for the RPA matrices. 
The symmetry is violated
because we need to superimpose terms with different $ \Gamma $ .\\
The implication of time-reversal invariance and the resulting
structure of the RPA matrix is a symmetry in the spectrum
of eigenvalues. We find that in this case the eigenvalues occur
in pairs such that for every $ \omega_{\Delta} $ there is an
adjoint $ - \omega_{\Delta}^{*} $. This way we are guaranteed to obtain 
as many positive as negative eigenvalues, the negative ones are then
discarded as unphysical. \\
On a nonscalar ground state this symmetry
is broken. 
In part II of this paper \cite{HGLM96} we will find that for the cases
under investigation the spectrum is still approximately
symmetric as long as the eigenvalues are real. \\
\subsection{The Basis States}
A genuine feature of the tensor-RPA approach is the overcompleteness
of the space of basis states defined by Eq.\ (\ref{rpaop}).
A basis state can be labelled by its $ph$ quantum numbers $ (mi,
\Gamma_k) $, where $m$ and $i$ denote the quantum numbers of
the shells in which particle and hole are created and $ \Gamma_k $
defines the $J/T$ quantum numbers the $ph$ pair is coupled to.
A novel feature of the tensor RPA is the fact that
the $ph$ space for fixed quantum numbers $ \Gamma_{\Delta} $
of the excited state 
contains states which are degenerate in energy
but carry different quantum numbers $ \Gamma_k $.
We will distinguish four different classes of basis states:
\begin{itemize}
\item[A.]
States with a hole in the Dirac sea and a particle in an empty shell.
\item[B.]
States with a hole in the Dirac sea and a particle in the valence shell.
\item[C.]
States with a hole in the valence shell and a particle in an
empty shell.
\item[D.]
States with a hole and a particle in the valence shell. 
\end{itemize}
For the moment we do not consider the states of class D,
since they require a modified RPA scheme.
We will come back to this issue later. \\
The metric tensor $N$ contains the mutual scalar products of the
basis states, so in principle we can answer all questions
concerning linear dependence and orthogonality by computing
the relevant matrix elements of $N$. \\
One can easily show that the states of class A are mutually orthogonal
and normalized, just like in the usual RPA.
The states of class B and C do not share that
feature, but we can derive some general statements about
the number of linearly independent states in each case. \\
We define the degeneracy $d$ of a $ph$ pair as the number
of allowed $J/T$ values it can be coupled to in order to
excite the ground state to a state
with quantum numbers $ \Gamma_{\Delta} = (J_{\Delta},T_{\Delta}) $. \\
If we restrict ourselves to
class B(C) we will always have a particle(hole) 
in the valence shell, the corresponding hole(particle), denoted by $r$,
being in the Dirac sea (in an empty shell).
In case of the nucleon we obtain for the states of class B and C
\be
d= \left\{ \begin{array}{ccl} 1 & : & j_r = J_{\Delta} \pm 1, \;
t_r = T_{\Delta} \pm 1 \\
2 & : & \left\{ \begin{array}{c} j_r = J_{\Delta}, \; t_r = T_{\Delta}
\pm 1 \\ j_r = J_{\Delta} \pm 1, \; t_r = T_{\Delta}
\end{array} \right. \\
4 & : & j_r = J_{\Delta}, \; t_r = T_{\Delta} . \end{array} \right.
\ee
The number of linearly independent basis states for given quantum
numbers $(mi)$, which we denote by $d_r $,
will be less than or equal to $d$.
In general the basis states are of the form
\be
| (\Gamma_k )_{mi} \rangle =
\left( (a_{m}^{\dagger} \times a_{\bar{i}} )^{\Gamma_k } \times 
| \Gamma_N \rangle \right)^{ \Gamma_{\Delta}}.
\label{basis}
\ee
The ansatz for the RPA operator also contains 
time reversed $ph$ operators, but the metric tensor $N$ can be
written in terms of mutual scalar products of the states in Eq.\ 
(\ref{basis}) only. 
We rewrite this expression by making use of the Racah-recoupling
theorem, but we have to make a distinction between the states B and
C. \\
For the B-states (i.\ e.\ states of class B) we obtain
\begin{eqnarray}
\lefteqn{\left( (a_{m}^{\dagger} \times a_{\bar{i}} )^{\Gamma_k } 
\times | \Gamma_N \rangle
\right)^{\Gamma_{\Delta}}} \nonumber \\ 
& & = (-1)^{ \Gamma_m + \Gamma_i - \Gamma_k +1 }
\sum_{ \Gamma} \hat{\Gamma}_k  \hat{\Gamma} \; W ( \Gamma_m \Gamma_i
\Gamma_{\Delta} \Gamma_N, \Gamma_k \Gamma ) 
\left( a_{\bar{i}} \times ( a^{\dagger}_m
\times | \Gamma_N \rangle )^{\Gamma} \right)^{\Gamma_{\Delta}} .
\label{recoupl1}
\end{eqnarray}
Effectively this is a unitary transformation from states
$ | (\Gamma_k )_{mi} \rangle $ with a coupled $ph$ pair
to states $ | \Gamma_{4q} \rangle $ with 4 coupled valence quarks. \\
In a similar way the C-states are reexpressed in terms of
states with two coupled valence quarks. Here the situation is slightly more
complicated, because we need an explicit representation of the nucleon
wavefunction. It is not difficult to verify that
\begin{eqnarray}
| \Gamma_N \Gamma_z \rangle  & = &
\left( \frac{\sqrt{2}}{3} a_{(\Gamma_N,\Gamma_z)}^{1 \dagger}
( a^{2 \dagger}_{\Gamma_N} \times a^{3 \dagger}_{\Gamma_N} )^{(0,0)}
- a_{(\Gamma_N,\Gamma_z)}^{2 \dagger} 
( a^{1 \dagger}_{\Gamma_N} \times a^{3 \dagger}_{\Gamma_N} )^{(0,0)}
\right. \nonumber \\ & & \left.
+ a_{(\Gamma_N,\Gamma_z)}^{3 \dagger} 
( a^{1 \dagger}_{\Gamma_N} \times a^{2 \dagger}_{\Gamma_N} )^{(0,0)}
\right) | 0 \rangle. 
\label{nuclprime}
\end{eqnarray}
Only the $J/T$ and color quantum numbers have been displayed.
When we insert this expression into Eq.\ (\ref{basis}) we get a contribution
from each term of the sum. After recoupling, the contribution from,
e.\ g.\ the third term, reads
\begin{eqnarray}
| (\Gamma_k)_{mi}, 3 \rangle & \equiv & \sum_{\Gamma} \hat{\Gamma}
\hat{\Gamma}_k  W( \Gamma_m \Gamma_i 
\Gamma_{\Delta} \Gamma_N, \Gamma_k \Gamma ) \nonumber \\
& & \cdot
\left( a_m^{c \dagger} \times \left( a_{\bar{i}}^{c} \times
\left( a^{3 \dagger}_{\Gamma_N} \times ( a^{1 \dagger}_{\Gamma_N}
\times a^{2 \dagger}_{\Gamma_N} )^{(0,0)}
\right)^{\Gamma_N} \right)^{\Gamma} \right)^{\Gamma_{\Delta}} | 0 \rangle .
\end{eqnarray}
Corresponding relations are found for the other terms.
Bearing in mind that $i$ is a state in the valence shell, we can
simplify this expression by commuting $ a_{\bar{i}}^c $ to the right.
We obtain
\begin{eqnarray}
\lefteqn{ | (\Gamma_k)_{mi}, 3 \rangle  =  
\sum_{\Gamma} \hat{\Gamma} \hat{\Gamma}_k
W (\Gamma_m \Gamma_i \Gamma_{\Delta} \Gamma_N, \Gamma_k \Gamma )} \nonumber \\
& & \cdot \left( \frac{1}{2} \left( a_m^{1 \dagger} 
\times ( a_{\Gamma_N}^{2 \dagger}
\times a_{\Gamma_N}^{3 \dagger} )^{\Gamma} 
\right)^{\Gamma_{\Delta}} 
-\frac{1}{2} \left( a_m^{2 \dagger} 
\times ( a_{\Gamma_N}^{1 \dagger} \times
a_{\Gamma_N}^{3 \dagger} )^{\Gamma} \right)^{\Gamma_{\Delta}}
\right. \nonumber \\
& & \left. + \delta_{\Gamma,(0,0)} \left( a_m^{3 \dagger} 
\times ( a_{\Gamma_N}^{1 \dagger}
\times a_{\Gamma_N}^{2 \dagger} )^{\Gamma} \right)^{\Gamma_{\Delta}} \right)
| 0 \rangle .
\label{recoupl2}
\end{eqnarray}
The first and second term in Eq.\ (\ref{nuclprime}) give similar expressions.
This shows that we can express the C-states as linear combinations of
states $ | \Gamma_{2q} \rangle $ with two coupled valence quarks
and a third quark in a higher shell. \\
As mentioned before, the number of linearly independent B- and C-states is 
smaller than
the degeneracy $d$. The reason for this is the fact that the
states $ | \Gamma_{4q} \rangle $ and $ | \Gamma_{2q} \rangle $
do not all comply with the Pauli principle for any given $ \Gamma_{4q},
\Gamma_{2q} $.
By applying Young-tableaux techniques it is found that the value
$ (J$$=$$0$,$T$$=$$0) $ is not allowed 
for $ \Gamma_{4q} $. For $ \Gamma_{2q} $
the values $ (0,1)$ and $(1,0)$ are excluded.
The derivation of these results is presented in Appendix B. \\
For given $ph$ quantum numbers $(mi)$ the forbidden states do
not in any case contribute to the right hand side of Eq.\ (\ref{recoupl1}) or
(\ref{recoupl2}). Due to the triangle selection rules some
of the $6j$-symbols vanish for certain values $ \Gamma = \Gamma' $. 
For this reason the states with the corresponding
quantum numbers do not
appear in the sum over $ \Gamma $. If, however, a $6j$-symbol
multiplying a forbidden state is nonzero the number of linearly
independent states $ d_r $ is lowered by one, since
the forbidden state has zero norm. 
An analysis of the triangle selection rules for the $6j$-symbols
reveals a simple relation between $d_r $ and the degeneracy $d$.
For the B-states we obtain
\be
d_r = \left\{ \begin{array}{ccl} 1 & : & d=1 \\
2 & : & d=2 \\ 3 & : & d=4 \; \; ,\end{array} \right. 
\ee
whereas for the C-states
\be
d_r = \left\{ \begin{array}{ccl} 1 & : & d=1 \\
1 & : & d=2 \\ 2 & : & d=4 \; \; .\end{array} \right. 
\ee
In a practical calculation one would have to reduce the dimension
of the naive $ph$ basis to $d_r $ in each degenerate subspace.
The reduced basis need not to be orthogonal, because nonorthogonal
states are taken care of by the metric tensor $N$, which contains
the mutual scalar products. Alternatively, a Schmidt-orthogonalization
procedure might be used. \\
Bearing in mind this reduction scheme the basis states of Eq.\ (\ref{basis})
completely span the space of physically allowed
$ph$ excitations on the nucleon.
They are color singlet states which belong to
$ SU(2)_J \otimes SU(2)_T $ multiplets and allow us to treat excitations of
the valence quarks and excitations of the Dirac sea on the same
footing. \\ 
In the nonrelativistic limit the C-states span the space of spin-flavor
wavefunctions one encounters in the nonrelativistic quark model.
There one discards certain spin-flavor representations from the
very beginning, because they constitute spurious center of mass
excitations (e.\ g.\ the symmetric spin-flavor representation for
the lowest $ \left( \frac{1}{2}^- , \frac{1}{2} \right) $ state).
In contrast to that we include the spurious states in our basis
and rely on the fact that they decouple from the excitation spectrum
in a consistent RPA calculation. In a relativistic framework
this is the method of choice, especially when there are broken
symmetries other than the translational symmetry,  
like in the case of mesonic modes related to chiral symmetry. \\
\subsection{The Interaction Matrix Elements}
So far we have only discussed issues which do not depend on the
structure of the Hamiltonian $H$. We now focus our attention on
the interaction matrix elements and give two different prescriptions
to compute these.  \\
We have to compute reduced matrix elements of the double commutators
appearing in Eq.\ (\ref{rpamatr}). To do so, we insert the definition
of the reduced matrix elements according to the
Wigner-Eckart theorem.
In case of, e.\ g.\, $M^{(1)}$ we get 
\be
\langle \Gamma_N || [ A_{mi} ( \bar{\Gamma}_k ) , H , A_{nj}^{\dagger}
(\Gamma_l ) ]^{\Gamma} || \Gamma_N \rangle = \hat{\Gamma}_N
\frac{
\langle \Gamma_N \Gamma_z | [ A_{mi} 
( \bar{\Gamma}_k ) , H , A_{nj}^{\dagger}
(\Gamma_l ) ]^{\Gamma}_0 | \Gamma_N \Gamma_z \rangle}
{\langle \Gamma_N \Gamma \Gamma_z 0 | \Gamma_N \Gamma_z \rangle} .   
\ee 
This expression is obviously independent of the projection quantum
numbers $\Gamma_z$, but we have to make a choice for $\Gamma_z $ in order to
perform the calculation. In the following we will not always
display the dependence of all quantities on the projection
quantum numbers and assume a fixed value of $\Gamma_z $ in case of doubt.
We write the interaction in the general form
\be
H_{int} = \frac{1}{2} \sum_{P,Q,R,S} \langle P Q | v |R S \rangle
a^{\dagger}_P a^{\dagger}_Q  a_S a_R .
\label{ww}
\ee
The sum is taken over the self-consistent single particle states.
We use the convention that upper case letters denote labels
that include the color quantum number (e.\ g.\ $P=(p,c_p )$), 
whereas lower case letters
label all quantum numbers besides color. \\
Usually one splits the interaction into a mean field and a residual
part by making use of Wick's theorem. The residual part is
defined via a normal ordered product with respect to the ground state
\be
N(a^{\dagger}_P a^{\dagger}_Q  a_S a_R)_{ ( \Gamma_N, \Gamma_z ) }.
\ee
With the theorems for normal ordered products \cite{FW71} it is then
easy to compute the RPA interaction matrix elements. \\
However, the application of Wick's theorem is only possible when
we can define a set of quasiparticle operators that annihilate the
ground state. This is no longer true for the nucleon state Eq.\
(\ref{nucl}). 
Even the most general ground state we can write down in the
framework of our model
\be
| N' \rangle = \sum_{\Gamma_z} C_{\Gamma_z} |\Gamma_N, \Gamma_z \rangle 
\ee 
contains correlations beyond the level of Bogolyubov quasiparticles
\cite{RS80} for any choice of the coefficients $ C_{\Gamma_z}$. \\
To compute the interaction matrix elements one can still
assume normal ordering, thus defining an {\em approximation}
that neglects the explicit ground-state correlations
which are due to the coupling of the valence quarks.
This approximation is solely based on algebraic relations
for the normal ordering operator and allows us to express
the RPA interaction matrix elements by the one-body
density of the ground state. \\
By Eq.\ (\ref{ww}) the residual interaction part of the reduced
matrix element becomes
\begin{eqnarray}
\lefteqn{ D_R (mi,nj) \equiv \frac{1}{2} 
\sum_{PQRS} \langle P Q | v |R S \rangle
\frac{\hat{\Gamma}_N}{\langle \Gamma_N 
\Gamma \Gamma_z 0 | \Gamma_N \Gamma_z \rangle} } \nonumber \\
& & \cdot \frac{1}{3} \langle \Gamma_N, \Gamma_z |
\left[ a^{\dagger}_{ic} a_{mc} ,
N( a^{\dagger}_P a^{\dagger}_Q  a_S a_R )_{ ( \Gamma_N, \Gamma_z ) } ,
a_{n c'}^{\dagger} a_{j c'} \right] | \Gamma_N, \Gamma_z \rangle . 
\label{resint}
\end{eqnarray}
In this expression all sums needed for the tensor coupling have been
suppressed and only the term entering $M^{(1)}$ has been
displayed. We obtain the analogous expressions for $M^{(2)}, M^{(3)}$ and
$ M^{(4)} $ if we make suitable interchanges of the quantum numbers
$ m,i,n$ and $j$. 
For the evaluation of the normal ordered product we use the
theorem \cite{FW71}
\begin{eqnarray}
\lefteqn{ N (\hat{O}_1 \ldots \hat{O}_{n-1} \hat{O}_n ) \hat{O}_{n+1} = }
\nonumber \\
& & N ( \! \! \trivcontract{&\hat{O}&_1  
\ldots \hat{O}_{n} & \hat{O}&_{n+1} } )
+ \ldots + N (\trivcontract{ \hat{O}_1 \ldots &\hat{O}&_{n} 
& \hat{O}&_{n+1} } )+ N( \hat{O}_1 \ldots \hat{O}_n \hat{O}_{n+1} ),
\end{eqnarray}
where $
\trivcontract{ & \hat{O}&_i & \hat{O}&_j } = \langle \hat{O}_i \hat{O}_j
\rangle $ denotes a contraction
with respect to $ | \Gamma_N, \Gamma_z \rangle $. \\
The $ \hat{O}_i$ can be any of the creation or annihilation
operators appearing in Eq.\ (\ref{resint}).
By successive use of the above relation we can express the
double commutator in Eq.\ (\ref{resint}) by contractions
of two operators and arrive at
\be
D_R (mi,nj) = \frac{1}{3} \frac{\hat{\Gamma}_N}
{\langle \Gamma_N \Gamma \Gamma_z 0
| \Gamma_N \Gamma_z \rangle } \langle (mc) (jc') | \bar{v} |
(ic) (nc') \rangle 
\cdot [ \rho(i) - \rho(m) ] [\rho(j) - \rho(n)],
\label{approx}
\ee
where $\rho$ is the density matrix as defined
in Eq.\ (\ref{dens}).
In this approximation the modifications with respect to the
usual RPA expression are basically the weight factors that
multiply the $ph$ matrix element. 
These weight factors account for the fact that the valence shell
is not fully occupied. \\
To compute the mean-field part of the RPA matrix elements, we assume
that the one-body part of the Hamiltonian, which is the sum of
the relativistic kinetic energy and the mean-field potential Eq.\ 
(\ref{vmf}), has been diagonalized 
\be
H_{MF} = \sum_K \epsilon_K a_K^{\dagger} a_K . 
\ee
At this stage an average over the projection quantum numbers of the
nucleon is assumed to identify the mean-field Hamiltonian
with the expression discussed in the previous chapter.
The mean-field contribution to the reduced 
matrix elements is then defined as
\be
D_{MF} (mi,nj) \equiv  
\sum_{K} \epsilon_K
\frac{\hat{\Gamma}_N}{\langle \Gamma_N \Gamma \Gamma_z 0 
| \Gamma_N \Gamma_z \rangle}
\cdot \frac{1}{3} \langle \Gamma_N, \Gamma_z |
\left[ a^{\dagger}_{ic} a_{mc} ,
a^{\dagger}_K a_K ,
a_{n c'}^{\dagger} a_{j c'} \right] | \Gamma_N ,\Gamma_z \rangle , 
\ee 
which can be simplified to give
\be
D_{MF} (mi,nj) = \frac{\hat{\Gamma}_N}
{\langle \Gamma_N \Gamma \Gamma_z 0 | \Gamma_N \Gamma_z \rangle}
\delta_{mn} \delta_{ij} ( \epsilon_m - \epsilon_i ) \cdot
[ \rho(i) - \rho(m) ] .
\label{rpamf}
\ee 
Again we find an expression well known from the usual formulation
of the RPA multiplied by weight factors. \\
In order to correctly incorporate the effects that are related to details
of the ground-state structure, we have to go beyond the normal ordering
approximation. Among the basis states with lowest energy there will
be many B- and C-states, and we expect the lowest lying excitations
to have a sizeable admixture of these.
The B- and C-states are those which are sensitive to the
structure of the valence shell, as can be seen from 
the occupation factors in Eq.\ (\ref{approx}).
Therefore it is worthwhile to compute the exact interaction
matrix elements and compare them to Eq.\ (\ref{approx}). \\ 
For this purpose we do not a priori split the interaction in two
parts, but simply compute the reduced matrix element of the full
interaction
\begin{eqnarray}
\lefteqn{ D_{int} (mi,nj) \equiv \frac{1}{2} 
\sum_{PQRS} \langle P Q | v |R S \rangle
\frac{\hat{\Gamma}_N}{\langle \Gamma_N \Gamma \Gamma_z 0 
| \Gamma_N \Gamma_z \rangle} } \nonumber \\
& & \cdot \frac{1}{3} \langle \Gamma_N, \Gamma_z |
\left[ a^{\dagger}_{ic} a_{mc} ,
a^{\dagger}_P a^{\dagger}_Q  a_S a_R ,
a_{n c'}^{\dagger} a_{j c'} \right] | \Gamma_N, \Gamma_z \rangle .
\label{fullint} 
\end{eqnarray}
From this expression we subtract the contributions 
from the mean-field potential Eq.\ (\ref{vmf}) and {\em define} the
remaining terms to be the residual interaction matrix elements $ D_R $.
By definition, the mean-field part of the RPA matrix elements
is then the same as in the normal ordering approximation. \\
For the evaluation of the expectation value in Eq.\ 
(\ref{fullint}) we express the ground state
by Eq.\ (\ref{nucl}) and commute the annihilation operators
in $ \langle \Gamma_N , \Gamma_z | $ to the right until
they act on the vacuum state $ | 0 \rangle $.
We find that there are three different kinds of contributions
to $ D_R $
\be
D_R = D_R^{(1)} + D_R^{(2)} + D_R^{(3)}.
\ee
The first one consists of all terms of Eq.\ (\ref{approx}) except those
weighted with a product of two occupation factors $g$ 
\begin{eqnarray}
\lefteqn{ D_R^{(1)} (mi,nj) = 
\frac{1}{54} \frac{\hat{\Gamma}_N}{\langle \Gamma_N \Gamma \Gamma_z 0
| \Gamma_N \Gamma_z \rangle } \langle (mc) (jc') | \bar{v} |
(ic) (nc') \rangle } \nonumber \\
& &  \cdot \left[ g(j) \theta(-i) + g(i) \theta(-j) -g(m) \theta(-j)
-g(n) \theta(-i) \right. \nonumber \\
& & -g(j) \theta(-m) -g(i) \theta(-n) +g(m) \theta(-n) +g(n) \theta(-m) 
\nonumber \\
& & + \left. 18(\theta(-j) \theta(-i) -\theta(-j) \theta(-m)
-\theta(-i) \theta(-n) + \theta(-n) \theta(-m)) \right].
\end{eqnarray} 
The second term only gives a contribution for two $ph$ states with
either the same particle or the same hole quantum numbers
\begin{eqnarray}
\lefteqn{ D_R^{(2)} (mi,nj) = \frac{1}{648} \frac{\hat{\Gamma}_N}
{ \langle \Gamma_N \Gamma \Gamma_z 0
| \Gamma_N \Gamma_z \rangle } (g(i)-g(m)+g(j)-g(n)) } \nonumber \\
& & \cdot  \sum_k g(k) \left[ \delta_{mn} \;
\langle (jc) (kc') | \bar{v} | (ic) (kc') 
\rangle - \delta_{ij} \; \langle (mc) (kc') | \bar{v} | (nc) (kc') \rangle 
\right] .
\label{term2}
\end{eqnarray} 
The third term cannot be expressed in terms of the occupation
factors $g$, but contains the coupling coefficients $T$ explicitly
\begin{eqnarray}
\lefteqn{ D_R^{(3)} (mi,nj) = \frac{1}{216} 
\frac{\hat{\Gamma}_N}{\langle \Gamma_N \Gamma \Gamma_z 0 
| \Gamma_N \Gamma_z \rangle } } \nonumber \\ & & \cdot
\sum_{P,Q,R,S}  C_{pq,rs} 
\langle P Q | \bar{v} | R S \rangle   
\left[  \delta_{ij} \delta_{rn}
G_{pq,ms} + \delta_{mn} \delta_{qj} G_{rs,ip} \right. \nonumber \\ & &  
\left. + \delta_{ij} \delta_{qm} G_{rs,np}  
+ \delta_{mn} \delta_{ri} G_{pq,js} 
- 4 \delta_{sn} \delta_{qm} G_{ip,jr} 
- 4 \delta_{pj} \delta_{ri} G_{nq,ms} \right].
\label{term3}
\end{eqnarray}
The projection on color-singlet configurations is ensured by
\be
C_{pq,rs}= \sum_{ \stackrel{c,c'=1}{c \neq c' } }^3 \delta_{c_p c}
\delta_{c_q c'} ( \delta_{c_r c'} \delta_{c_s c} -
\delta_{c_r c} \delta_{c_s c'}) .
\ee
The color operator $C_{pq,rs}$ represents those contributions
from the two-body interaction $a^{\dagger}_P a^{\dagger}_Q a_S a_R$
where a creation and an annihilation operator are coupled
to an intermediate color-octet. 
Note the $+$ sign in front of
the exchange term in Eq.\ (\ref{term3}), which is due to the
antisymmetry of $C_{pq,rs}$ under exchange of $r$ and $s$. \\
The structure of the valence shell is reflected in a
matrix $G$, which is defined as
\be
G_{pq,rs}= \left\{ \begin{array}{lcl} 
\sum_{\Gamma_z } T_{\Gamma_{zp} \Gamma_{zq} \Gamma_z}
T_{\Gamma_{zr} \Gamma_{zs} \Gamma_z } & : & p,q,r,s \; \epsilon \; 
$valence shell$ \\
0 & : & $else$ \end{array} \right. .
\ee
Through terms $D_R^{(2)}$ and $D_R^{(3)}$ new types of matrix elements
are introduced to the RPA theory.
Eq.\ (\ref{term2}) describes the scattering of particles
and holes by the mean field of the valence quarks.
In addition to scattering processes (i.\ e.\ processes in which
particles or holes keep their identity) 
Eq.\ (\ref{term3}) also describes the transition of particles and holes
into valence states. 
Similar structures are found in QRPA, where $ph$ and 
$pp/hh$ matrix elements contribute. \\
\subsection{Description of $ (\frac{3}{2}^+ , \frac{3}{2} )$ states}
With the formalism presented so far it is not possible to study
excitations of the nucleon in the $ (\frac{3}{2}^+ , \frac{3}{2} )$
channel. The three valence quarks can as well be coupled to
the quantum numbers of the $ \Delta(1232) $, which can be achieved
by acting with a $ph$ operator of special structure on the
nucleon ground state
\be
| \Gamma_{\Delta} \rangle = ( A_{mi}^{\dagger} 
(\Gamma) \times | \Gamma_N \rangle )^{\Gamma_{\Delta}},
\label{recoupl}
\ee
where $ \Gamma=(1,1) $.
In this expression both the particle $m$ and the hole $i$ 
are in the valence shell. If the mean-field level of the theory
is defined in the sense of the previous section (with an averaged
ground-state density) such a state is clearly energetically
degenerate with the nucleon.
For the $ph$ operator in Eq.\ (\ref{recoupl}) we have
\be
A_{mi}^{\dagger} (\Gamma) = A_{mi} (\bar{\Gamma}).
\label{herm}
\ee
The adjoint of an RPA excitation operator including contributions
from the recoupling operator Eq.\ (\ref{recoupl}) will in general
not fulfill the relation
\be
O_{x \Gamma_k} | \Gamma_N \rangle \approx 0,
\label{annihil}
\ee 
since the $ph$ annihilation operators taken from Eq.\ (\ref{recoupl})
are of the same structure as the creation operators.
Hence, there is no reason to believe that the $Y$-Amplitudes
multiplying the recoupling contribution are small.
However, in the derivation of the RPA equation we have assumed
that Eq.\ \ref{annihil} is fulfilled. \\
From investigations of hyperfine splitting in the MIT-bag model
it is strongly suggested that the wavefunction of the $ \Delta(1232)$
has a dominant contribution from the recoupling configuration
\cite{DJJK75}. Thus an RPA approach in a $ph$ space without the state
Eq.\ (\ref{recoupl}) can at best account for the higher lying
resonances in the $ (\frac{3}{2}^+ , \frac{3}{2} )$ channel. \\
In the following derivation we neglect the tensor coupling
of operators and states in order to point out more clearly
the special features of the equations for the $ \Delta $-channel.
A straightforeward way to account for admixtures
of both types of configurations 
is to write down the full Hamiltonian
in a model space containing the recoupled ground state
as well as the nondegenerate $ph$ excitations.
For this purpose we write an excited state as
\be
|x \Gamma_{\Delta} \rangle = | \Delta_v \rangle
+ | \Delta_k \rangle,
\ee
where $x$ denote additional quantum numbers.
The projection on the recoupling configuration is given by
\be
| \Delta_v \rangle = Z_v | v \rangle,
\ee
where $ |v \rangle $ is the properly normalized state
of Eq.\ (\ref{recoupl}). The projection on the space of $ph$ states
(of nonzero energy) is given by
\be
| \Delta_{k} \rangle = \sum_{i} Z_{k_i} | k_i \rangle ,
\ee 
where
\be
| k_i \rangle =  \eta^{\dagger}_{k_i} | \Gamma_N
\rangle 
\ee
and $ \eta^{\dagger}_{k_i} $ are $ph$ operators including the
states of class A,B and C. Again we assume proper normalization
of the basis states. All quantum numbers, including angular
momentum and isospin, have been absorbed in the label $k_i $.
Clearly, the two subspaces are orthogonal, i.\ e.\ $ \langle k_i
| v \rangle = 0 \; \forall \: i $. Configuration mixing
is contained in the amplitudes $ Z_v, Z_{k_i} $, where
\be
| Z_v |^2 + \sum_i | Z_{k_i} |^2 = 1.
\ee
To proceed, we split off a residual interaction $V$ from the
Hamiltonian, in the same way as this was done in the previous section.
We assume that the mean-field part of the Hamiltonian is 
diagonal in the basis we have chosen. The projection of the
eigenvalue equation on the different subspaces is given by
\begin{eqnarray}
( H_{v v} - E) Z_{v} +
\sum_{i} V_{v k_i} Z_{k_i} & = & 0 \nonumber \\
\sum_{j} ( H_{k_i k_j} - \delta_{k_i k_j} E) 
Z_{k_j} + V_{k_i v} Z_{v} & = & 0 .
\label{project}
\end{eqnarray}
Clearly $ H_{k_i k_j} $ and $ V_{v k_i} $ are matrices according to
the dimensionality of the $ph$ space. The matrix elements
are in general divergent due to contributions from the
Dirac sea. A prescription to regularize these matrix elements
in agreement with the cut-off scheme for the mean-field
Hamiltonian has been given in \cite{Har96}. When we solve
the first equation for $ Z_v $ and insert the result into
the second, we obtain
\be
\sum_{j} (H_{k_i k_j} + \Sigma_{k_i k_j} (E) - 
\delta_{k_i k_j} E ) Z_{k_j} = 0 ,
\label{dispeqn}
\ee
where a self-energy
\be
\Sigma_{k_i k_j} = - V_{k_i v} G_{v} V_{v k_j}
\ee
accounts for the coupling to the valence configuration.
$ G_v $ is the projection of the many-body propagator
and is given by
\be
G_v = \frac{1}{H_{vv} - E} .
\ee
Thus in the $ph$ channel we obtain an equation where
the residual interaction is supplemented by a self-energy term,
compared to the conventional formulation without coupling
to the valence configuration.
We immediately find that the self energy will give
a considerable contribution to Eq.\ (\ref{dispeqn})
for energies close to the pole value $ H_{\alpha \alpha} $.
Since Eq.\ (\ref{project}) is a diagonalization problem
for a hermitian operator, the lowest eigenvalue will be located
below $ H_{\alpha \alpha} $. \\ 
Denoting the energy-expectation value of
the nucleon ground state Eq.\ (\ref{nucl}) by $ E_0 $, we 
obtain for the $N$-$\Delta$ mass splitting
\be
\Delta E_{N \Delta} < H_{\alpha \alpha} -E_0.
\label{massplit}
\ee 
This quantity must be compared to the phenomenological value for the
mass splitting of about $300 MeV$. In the last section it was argued
that $ H_{\alpha \alpha} -E_0 $ should already be a rather small quantity
for reasons of consistency with the assumptions entering
the derivation of the RPA equation. Thus it might be difficult to
reproduce the observed value for the mass splitting on the
level of Eq.\ (\ref{massplit}). \\ 
A solution to this problem
could be provided when we treat the nucleon ground state on the
same basis as the recoupling configuration. Due to 
the averaging of the ground-state density over projection
quantum numbers in the course which information about the
detailed structure of the nucleon is lost we can lower the
nucleon energy $E_0$ by infinitesimal $1p-1h$ admixtures.
In the usual Hartree-Fock description such admixtures are
excluded by the self-consistency condition. The derivation
presented in this section effectively takes into account
$1p-1h$ admixtures in the $ (\frac{3}{2}^+ , \frac{3}{2} )$ channel
to the recoupling configuration $ | \alpha \rangle $.
In a similar way we have to account for $1p-1h$ admixtures to
the nucleon in the $ (\frac{1}{2}^+ , \frac{1}{2} )$ channel.
This is required by reasons of consistency when comparing the
$ (\frac{1}{2}^+ , \frac{1}{2} )$ and the $ (\frac{3}{2}^+ , \frac{3}{2} )$
channel. By such admixtures the energy of the nucleon will be
lowered, thus the mass splitting is increased with respect to
the value obtained from configuration mixing
in the $ \Delta$-channel. \\
In addition to that we have to take into account another
important point.
So far the spectrum of $ (\frac{3}{2}^+ , \frac{3}{2} )$ states
was derived from a diagonalization problem of the full
Hamiltonian in a model space. It is important to note that
the usual RPA scheme is not equivalent to such a diagonalization problem.
However, even when we choose excitation operators
$ \eta^{\dagger}_{k_i} $ of RPA form in Eq.\ (\ref{rpaop})
we still obtain a diagonalization problem, since the
contribution of the time reversed operators vanishes.
In the subspace of $ph$ states we recover the well known
RPA structure when we replace the operator products in
Eq.\ (\ref{project}) by double commutators
\be
\langle \Gamma_N | \eta_{k_i} H \eta^{\dagger}_{k_j} | \Gamma_N \rangle
= \langle \Gamma_N | [ \eta_{k_i}, H , \eta^{\dagger}_{k_j} ] 
| \Gamma_N \rangle + E_0 \langle \Gamma_N | [ \eta_{k_i}, 
\eta^{\dagger}_{k_j} ] | \Gamma_N \rangle .
\label{substi1}
\ee
Note that within the space of $ph$ operators we have
\be
\eta_{k_i} | \Gamma_N \rangle = 0,
\label{annihil2}
\ee
which is an approximate identity when RPA operators are acting
on a HF ground state.
This justifies the replacement in Eq.\ (\ref{substi1}).
Note that we are not allowed to introduce double commutators
in the full space of excitation operators, since the recoupling
operator does not annihilate the ground state.
Abbreviating the recoupling operator of Eq.\ (\ref{recoupl})
with $ \eta^{\dagger}_v $, we are, however, allowed to
subtract the ground state energy   
\be
\langle \Gamma_N | \eta_v H \eta^{\dagger}_v | \Gamma_N \rangle
= \langle \Gamma_N | \eta_v [H, \eta^{\dagger}_v ]| \Gamma_N \rangle
+ E_0 \langle \Gamma_N | \eta_v \eta^{\dagger}_v | \Gamma_N \rangle .
\label{substi2}
\ee
Together with Eq.\ (\ref{substi1}) we thus obtain
an eigenvalue equation for the excitation energy 
$ \omega_{\Delta} = E_{\Delta} - E_0 $.
It is easy to show that the replacements of Eq.\ (\ref{substi1})
and (\ref{substi2}) preserve the symmetry of the eigenvalue problem
(i.\ e.\ hermiticity) if Eq.\ (\ref{annihil2}) is fulfilled
exactly, which
is the case on Tamm-Dancoff level.
For excitation operators of RPA type we obtain, as usual,
a diagonalization problem for a nonhermitian matrix. Wheras
in case of a scalar system
such an RPA problem can in general be reformulated
as an eigenvalue equation for a hermitian matrix \cite{RS80},
the tensor RPA scheme does not allow for such a reformulation.
This fact is connected to the asymmetry of the spectrum
under inversion of the energy axis. \\
In such a more general formulation Eq.\ (\ref{massplit})
looses its validity. The upper bound for the $N$-$\Delta$ mass splitting
is based on the Rayleigh-Ritz variational principle
for hermitian matrices. Thus for the description of
$ (\frac{3}{2}^+ , \frac{3}{2} )$ states it seems to be
crucial to go beyond the Tamm-Dancoff level and
introduce commutators instead of operator products. \\
On RPA level there is no a priori estimate for the mass
of the $ \Delta(1232) $ which stands in contradiction
to the consistency requirements set by the Hartree-Fock theory.
Only in a numerical calculation for a specific model
it can be decided if such a RPA scheme allows for a
realistic description of the $ (\frac{3}{2}^+ , \frac{3}{2} )$ 
spectrum.
\section{Summary, Discussion and Conclusion}
The theory presented in this paper is intended to provide a consistent
approach to many--body effects in the ground state and excited states of
the nucleon. By using a fieldtheoretical formulation essential features
of low--energy hadron physics are accounted for.
The theory is based on an
effective two--body quark--quark interaction which must be
determined empirically. 
The approach describes baryon spectroscopy by an expansion
in terms of many-body correlation functions, where at the
present stage RPA contributions were investigated. \\
The nucleon is considered as a system of three valence quarks which 
interact with the vacuum. The ground state of the full system,
including valence and vacuum parts, was described with Hartree--Fock
methods which clearly represents a nonperturbative approach.
By introducing a symmetry conserving regularization of the
mean-field potential a self-consistent RPA scheme including
excitations from the Dirac sea was defined. \\ 
A central issue of the model is that states of
good angular momentum $J$ and isospin $T$ are used. 
This we consider as an
important advantage over other approaches which are based on a grand
spin description, as e.\ g.\ the hedgehog soliton \cite{ARW95,CGGP95}. 
However, it was
discussed that this leads to 
mean--field self--energies violating rotational and isospin invariance. 
Although many--body theory provides in
principle methods to treat such a problem, e.\ g.\ in nuclear structure
physics \cite{RS80}, we have chosen a somewhat simplified approach by
averaging over projection quantum numbers. This led to a one--body
density matrix which is the same for the $(\frac{1}{2}^+,\frac{1}{2})$
nucleon and the 
$(\frac{3}{2}^+,\frac{3}{2})$ $\Delta$-like configuration.
A mean-field Hamiltonian defined with the averaged density
allows to extract single particle states with good $J/T$. From
these a nucleon state can be built which contains important
details of the many-body structure related to angular momentum
and isospin coupling. \\
Excited states of the nucleon were described by RPA--methods. Since we
are dealing with a nonscalar ground state the
theory must account for the mixing of $ph$ configurations with
different $J/T$ quantum numbers in an excited state.
Appropriate
methods are provided by the tensor RPA which originally was introduced
in nuclear structure physics by Rowe et al. \cite{RN75}. 
Although technically rather involved, the tensor RPA method relies
on a similar representation of excited states as collective
$1p-1h$ configurations on a mean-field ground state as
the standard RPA scheme.
In the present case,
this includes transitions from the valence sector and the Dirac--sea
into positive energy states. The application of RPA theory to the
vacuum sector is well established for the NJL-model \cite{KLVW90}.
As an extension of such techniques we have made a 
first attempt to describe a baryon with 
good J/T quantum numbers in the ground state and in the
excited states. \\
The description of excited states provides additional information on the
residual interaction beyond those properties which contribute to the
ground state. It was pointed out that the regularization of
the Hamiltonian led to a modification of the residual interaction.
This reflects the fact that a projection onto a model space is used
which includes vacuum states only up to a
certain cut--off energy. For practical purposes it might be
preferable to use a parametrization of the residual interaction
and account for the modifications implied by the cut-off scheme
by a readjustment of the parameters. \\
It was also pointed out that the RPA theory allows to
obtain information on the ground-state structure
beyond mean-field level. On the one hand there
will be $1p-1h$ admixtures which are due to the
averaging prescription for the ground-state density.
On the other hand an RPA ansatz includes implicit
$2p-2h$ correlations. 
An improved ground-state wavefunction including such
contributions will allow to recover the coupling
to pion degrees of freedom which are excluded on
mean-field level for reasons of symmetry.
Furthermore ground-state correlations will lead
to a depolarization of the valence-quark core
and allow to study the spin transfer to, e.\ g.\,
excitations of the Dirac sea.
Such admixtures have already been
investigated in \cite{GLM96} using a somewhat simplified
RPA scheme.
\section*{Appendix A}
In this appendix we derive an operator equation for the effective $ph$
interaction which has to be used in the Dirac-RPA equation. \\
By definition, the mean-field Hamiltonian depends on $\rho $
only via the the regularized density $ \rho'$
\begin{equation}
\frac{\partial}{\partial \rho_{k k'}} H_{MF}' =
\frac{\partial}{\partial \rho_{k k'}} H_{MF} (\rho') =
\left( \frac{\partial}{\partial \rho'_{l l'}} H_{MF} \right)
\frac{\partial \rho'_{l l'}}{\partial \rho_{k k'}}.
\end{equation}
We have adopted the convention to sum over repeated indices.
The regularized density depends on $ \rho $ also via
the regularization function $ R $
\begin{equation}
\frac{\partial \rho'_{l l'}}{\partial \rho_{k k'}} =
\frac{\partial R_{l i}}{\partial \rho_{k k'} } \rho_{i i'} R_{i' l'}
+ R_{l i} \diff{ \rho_{i i'}}{ \rho_{k k'}} R_{i' l'} 
+ R_{l i} \rho_{i i'} \diff{ R_{i' l'}}{ \rho_{k k'} } .
\end{equation}
We arrive at
\begin{eqnarray}
\lefteqn{\frac{\partial }{\partial \rho_{k k'}} H_{MF}' =
\left( \frac{ \partial}{ \partial \rho_{l l'}'} H_{MF} \right)
R_{lk} R_{lk'}} \nonumber \\
& & + \left( \frac{\partial}{\partial \rho_{l l'}'} H_{MF} \right)
\rho_{i i'} \diff{ \left( H_{MF}' \right)_{m m'} }{ \rho_{k k'} }
\left[ \diff{ R_{li}}{ \left( H_{MF}' \right)_{m m'} }R_{i' l'}
+ R_{li} \diff{ R_{i' l'}}{ \left( H_{MF}' \right)_{m m'}} \right] .
\label{impl}
\end{eqnarray}
The effective interaction $ \partial H_{MF}' / \partial \rho $
appears on both sides of the equation.
Expression (\ref{impl}) has to be evaluated at $\rho_0 $, the stationary
solution of the TDHF equation. For the remaining parts of this
section we adopt the convention that single particle labels refer
to the basis in which $ \rho_0 $ is diagonal. \\ 
Denoting the eigenvalues of $ H_{MF}' $ by $ \epsilon_i $, we
find that the first term on the right hand side of Eq.\ (\ref{impl})
is given by 
\be
\left. \left( \frac{ \partial}{ \partial \rho_{k k'}'}
H_{MF} \right) \right|_{\rho_0} R[ \epsilon_k ] R[ \epsilon_{k'} ]
\ee
where all density-dependent quantities have been evaluated
at $ \rho_0 $.
In the self-consistent basis the density matrix acquires the form
\be
\left( \rho_0 \right)_{k k'} = \delta_{k k'} \theta( k_F -k),
\ee
for states normalized to unity.
The symbolic $\theta$-function indicates that $k$ has to be
below the Fermi surface which is defined by $k_F$. 
In an infinite system $k_F$ denotes the Fermi momentum. \\
To evaluate the second term in Eq.\ (\ref{impl}), we write
the regularization function as
\be
R[x] = \sum_{n=0}^{\infty} a_n x^{n}.
\ee
For an arbitrary operator $A$ we have
\begin{eqnarray}
& & \diff{R_{li}}{ A_{m m'}}= 
\sum_{n=0}^{\infty} a_n \left(
\diff{ A_{li_1 } }{ A_{m m'} } 
A_{i_1 i_2} \ldots A_{i_{n-1} i}  \right. \nonumber \\ 
& & \left. + A_{l i_1 } 
\diff{ A_{i_1 i_2} } { A_{m m'} } \ldots 
A_{i_{n-1} i} + \ldots + A_{l i_1} 
A_{i_1 i_2} \ldots \diff{ A_{i_{n-1} i}}
{ A_{m m'} } \right),
\end{eqnarray}
where the Einstein convention for repeated indices is used.
Evaluating this expression for $ A= H_{MF}' $ gives
\be
\left. \diff{R_{li}}{( H_{MF}')_{m m'} } \right|_{\rho_0} =
\delta_{l m} \delta_{i m'} \sum_{n=0}^{\infty} a_n
( \epsilon_{m'}^{n-1} + \epsilon_m \epsilon_{m'}^{n-2} + \ldots
+ \epsilon_{m}^{n-2} \epsilon_{m'} + \epsilon_{m}^{n-1} ).
\ee
The expression in brackets can be cast into the form of a finite
geometrical series. Then the derivative of the cut-off function
simplifies to 
\be
\left. \diff{R_{li}}{( H_{MF}')_{m m'} } \right|_{\rho_0} =
\frac{ \delta_{lm} \delta_{i m'}}{ \epsilon_{m'} - \epsilon_{m} }
\left( \sum_{n=0}^{\infty} a_n \epsilon_{m'}^{n}
- \sum_{n=0}^{\infty} a_n \epsilon_{m}^{n} \right).
\ee
On the right hand side we recover the defining power series
for the cut-off function $R$ and obtain
\be
\left. \diff{R_{li}}{( H_{MF}')_{m m'} } \right|_{\rho_0} =
\delta_{lm} \delta_{i m'} \frac{ R[ \epsilon_{m'}] - R[ \epsilon_{m} ]}
{ \epsilon_{m'} - \epsilon_{m} }.
\ee
With that we finally obtain an operator equation from which 
the effective interaction can be determined
\begin{eqnarray}
\lefteqn{ \left. \frac{ \partial}{ \partial \rho_{k k'} } 
H_{MF}' \right|_{\rho_0} =
\diff{ H_{MF}}{ \rho_{k k'}' } R[\epsilon_k] R[\epsilon_{k'} ] }
\\
& & + \left. \diff{H_{MF}}{\rho_{l l'}'} \right|_{\rho_0}
\left. \diff{ \left( H_{MF}' \right)_{l l'} }{ \rho_{k k'}} 
\right|_{\rho_0} \frac{R[\epsilon_{l'} ]- R[\epsilon_{l}] }{ \epsilon_{l'}
- \epsilon_l } \left( \theta (k_F -l) R[ \epsilon_l]
+ \theta (k_F -l') R[ \epsilon_{l'}] \right).
\nonumber
\end{eqnarray}
\section*{Appendix B}
The structure of $3$-quark wavefunctions has been discussed
in detail in the literature, e.\ g.\ in the book
by Close \cite{C79}.
In this appendix we investigate the group representations which
can build from the $2$- and $4$-quark wavefunctions we encounter
when we recouple the RPA basis states. \\
First we discuss the B-states, which contain a hole in the Dirac sea
and four quarks in the valence shell. \\
A single particle state is characterized by its quantum numbers
in Dirac, isospin and color space. We assume that the valence shell
is a $ | \kappa | = 1 $ state, 
so that the Dirac-space part of the 
wavefunction is uniquely determined by specifying the angular-momentum
projection quantum number. We can simply imagine having spin
$ \frac{1}{2}$ particles with additional isospin and color degrees
of freedom, so we will label the Dirac-space representations
with a subscript $J$, denoting the total angular momentum. 
The isospin and color 
representations carry subscripts $T$ and $C$, respectively. \\
We have to construct irreducible representations of the group
$ SU(2)_J \otimes SU(2)_T \otimes SU(3)_C $. In order to be allowed by
the Pauli principle, these states have to form antisymmetric
representations of the symmetric group. The space of allowed
$4$-quark wavefunctions is further restricted to those combinations
which can be reached by coupling one quark to a $(J$=$\frac{1}{2}$,
$T$=$\frac{1}{2})$ color singlet $3$-quark state. \\
According to the rules for outer products of Young tableaux \cite{L78,H62}
the only relevant $SU(2)$ diagrams are $(3,1)$ and $(2,2)$
with dimensionalities
\be
d_{SU(2)} = \left\{ \begin{array}{lcr} 3 & : & (3,1) \\
1 & : & (2^2) \end{array} \right. \; 
\ee 
When we take the inner product of a $J$ with a $T$ representation
we can classify the resulting representation by a $ SU(4)_{JT} $
Young tableaux. We could now deduce the $ SU(2)_J \otimes SU(2)_T $
content of the $SU(4)$ representations by using the 
Clebsch-Gordan coefficients of the symmetric group \cite{H62}. In many cases
this is not neccessary, since one often can conclude on which
product representations of lower dimensional unitary groups
are contained in a Young tableaux by considering the symmetry
of the diagrams and matching the dimensionalities of the representations. \\
One finds that the product representations of the relevant $J$ and
$T$ diagrams are contained in $SU(4)_{JT}$ diagrams, according to
\be
\begin{array}{l|l}
SU(2)_J \times SU(2)_T & SU(4)_{JT} \\
\hline
(3,1) \times (3,1) & (4),(3,1),(2^2),(2,1^2) \\
(3,1) \times (2^2) & (3,1),(2,1^2) \\
(2^2) \times (3,1) & (3,1),(2,1^2) \\
(2^2) \times (2^2) & (4),(2^2),(1^4) \hspace{2cm} .
\end{array}
\ee 
The $SU(3)_C$ representation is fixed to be $(2,1^2)$, since the
three valence quarks of the nucleon form a color singlet. 
Coupling this $C$-representation
on the $SU(4)$ diagrams one can embed the resulting product representations
in multiplets of the group $SU(12)_{JTC}$, according to
\be
\begin{array}{l|l}
SU(4)_{JT} \times SU(3)_C & SU(12)_{JTC} \\
\hline
(4) \times (2,1^2) & (2,1^2) \\
(3,1) \times (2,1^2) & (1^4),(2,1^2),(2^2),(3,1) \\
(2^2) \times (2,1^2) & (3,1),(2,1^2) \\
(2,1^2) \times (2,1^2) & (4),(3,1),(2^2),(2,1^2) \\
(1^4) \times (2,1^2) & (3,1) \hspace{2cm} .
\end{array}
\ee
The $SU(12)$ representation is restricted to the completely
antisymmetric $(1^4)$. Therefore the only allowed $SU(4)$ diagram
is $(3,1)$. From the spin-singlet isospin-singlet diagrams we
cannot construct this $SU(4)$ representation, so a configuration
of four valence quarks with $(J$=$0$,$T$=$0)$ is excluded by the
Pauli principle. \\
For the states of class C we can apply quite similar arguments based on
a configuration of two valence quarks. To obtain the possible
$2$-quark Young tableaux, we again label the Dirac space diagrams with
a subscript $J$, since all other quantum numbers are the same.
Then the dimensionalities of the Dirac and isospin diagrams are
\be
d_{SU(2)}= \left\{ \begin{array}{lcr} 3 & : & (2) \\
1 & : & (1^2) \end{array} \right. \; .
\ee
These representations can be coupled according to
\be
\begin{array}{l|l}
SU(2)_J \times SU(2)_T & SU(4)_{JT} \\
\hline
(2) \times (2) & (2) \\
(2) \times (1^2) & (1^2) \\
(1^2) \times (2) & (1^2) \\
(1^2) \times (1^2) & (2) \hspace{2cm} .
\end{array}
\ee
When we take the outer product of the $2$-quark $SU(4)$ diagrams with a third
particle in a higher shell, we obtain
\begin{eqnarray}
& & (2) \otimes (1) = (3) + (2,1) \nonumber \\
& & (1^2) \otimes (1) = (2,1) + (1^3) \; .
\end{eqnarray}
In color space the three quarks form a completely antisymmetric
representation, so the residual wavefunction has to be
symmetric under exchange of particles. Only the symmetric
$SU(4)$ representation for two quarks can be coupled to
a symmetric $3$-quark representation, so the $(J$=$0$,$T$=$1)$ and
the $(J$=$1$,$T$=$0)$ $2$-quark configurations are forbidden
by the Pauli principle.

\end{document}